\documentclass[journal]{IEEEtran}
\IEEEoverridecommandlockouts
\usepackage{amsmath,amssymb,amsfonts}
\usepackage{algorithm}
\usepackage{algorithmic}
\usepackage{amsmath}
\usepackage{amssymb}
\usepackage{array}
\usepackage{booktabs}
\usepackage{cite}
\usepackage{caption}
\usepackage{color}
\usepackage{enumitem}
\usepackage[T1]{fontenc}
\usepackage{graphicx}
\usepackage{harveyballs}
\usepackage{makecell}
\usepackage{tabularx}
\usepackage{multirow}
\usepackage{stfloats}
\usepackage{subcaption}
\usepackage{textcomp}
\usepackage{tabu}
\usepackage{url}
\usepackage{listings}
\usepackage{ragged2e}
\usepackage{xurl}

\usepackage{graphicx}  
\usepackage{adjustbox} 
\usepackage{booktabs}  

\usepackage{xcolor}

\newcommand{\revRonetwo}[1]{#1}

\def\BibTeX{{\rm B\kern-.05em{\sc i\kern-.025em b}\kern-.08em
    T\kern-.1667em\lower.7ex\hbox{E}\kern-.125emX}}

\begin{document}

\title{Twinning for Space-Air-Ground-Sea Integrated Networks: Beyond Conventional Digital Twin Towards Goal-Oriented Semantic Twin
\thanks{
	This work was supported in part by the National Natural Science Foundation of China under Grant nos. 62571159, 61871147, in part by the National Key Research and Development Program of China under Grant no. 2020YFB1806403, and in part by the Major Key Project of PCL under Grant no. PCL2024A01.  (Yifei Qiu, Tianle Liao and Xin Jin contributed equally to this work.) (Corresponding author: Shaohua Wu.) 
	
	Y. Qiu, T. Liao and X. Jin are with the School of Information Science and Technology, Harbin Institute of Technology (Shenzhen), Shenzhen 518055, China (e-mail: qiuyifei@stu.hit.edu.cn; 25B952014@stu.hit.edu.cn; 24S052022@stu.hit.edu.cn).
	
	 Q. Zhang and S. Wu  are with the Guangdong Provincial Key Laboratory of Aerospace Communication and Networking Technology, Harbin Institute of Technology (Shenzhen), Shenzhen 518055, China, and also with the Pengcheng Laboratory, Shenzhen 518055, China (e-mail: zqy@hit.edu.cn; hitwush@hit.edu.cn).

 Copyright (c) 2026 IEEE. Personal use of this material is permitted. However, permission to use this material for any other purposes must be obtained from the IEEE by sending a request to pubs-permissions@ieee.org. 
}

}

\author{\IEEEauthorblockN{Yifei~Qiu, 
    Tianle~Liao,
    Xin~Jin,  
	and Qinyu~Zhang,~\IEEEmembership{Senior Member,~IEEE,}
    	Shaohua~Wu,~\IEEEmembership{Senior Member,~IEEE}\\
}
}

\maketitle

\begin{abstract}
The space-air-ground-sea integrated network (SAGSIN) has emerged as a cornerstone of 6G systems, establishing a unified global architecture by integrating multi-domain network resources. Motivated by the demand for real-time situational awareness and intelligent operational maintenance, digital twin (DT) technology is regarded as a promising solution, owing to its capability to create virtual replicas and emulate physical system behaviors. However, the large scale, high dynamics, and weak connectivity of SAGSIN pose severe challenges to DT deployment, as conventional DT paradigms struggle to address excessive computational overhead, model synchronization latency, and cross-domain platform heterogeneity. To address these limitations, this survey proposes a novel twinning framework: the goal-oriented semantic twin (GOST). GOST prioritizes utility over fidelity, leveraging semantic technologies to exploit intrinsic relationships among entities and adhering to goal-oriented principles to construct lightweight, task-specific representations. This paper systematically elucidates the GOST framework through three layers: knowledge-based semantics, data-driven semantics, and goal-oriented principles. Furthermore, we summarize the core technologies underpinning GOST and formulate a multi-dimensional evaluation methodology. We also present a case study on collaborative tracking in remote satellite-UAV networks, demonstrating that GOST outperforms conventional DTs in terms of perceptual data timeliness and tracking accuracy. Finally, we chart future research directions, positioning GOST as a promising twinning paradigm for SAGSIN development.

\end{abstract}

\begin{IEEEkeywords}
Space-air-ground-sea integrated network, digital twin, goal-oriented semantic twin, perception-communication-computing-actuation (PCCA), deep learning, distributed computing.
\end{IEEEkeywords}

\section{Introduction} \label{I}

 \subsection{Background and Motivation}

\begin{figure*}[htbp]
    \centering
    \includegraphics[width=0.95\textwidth]{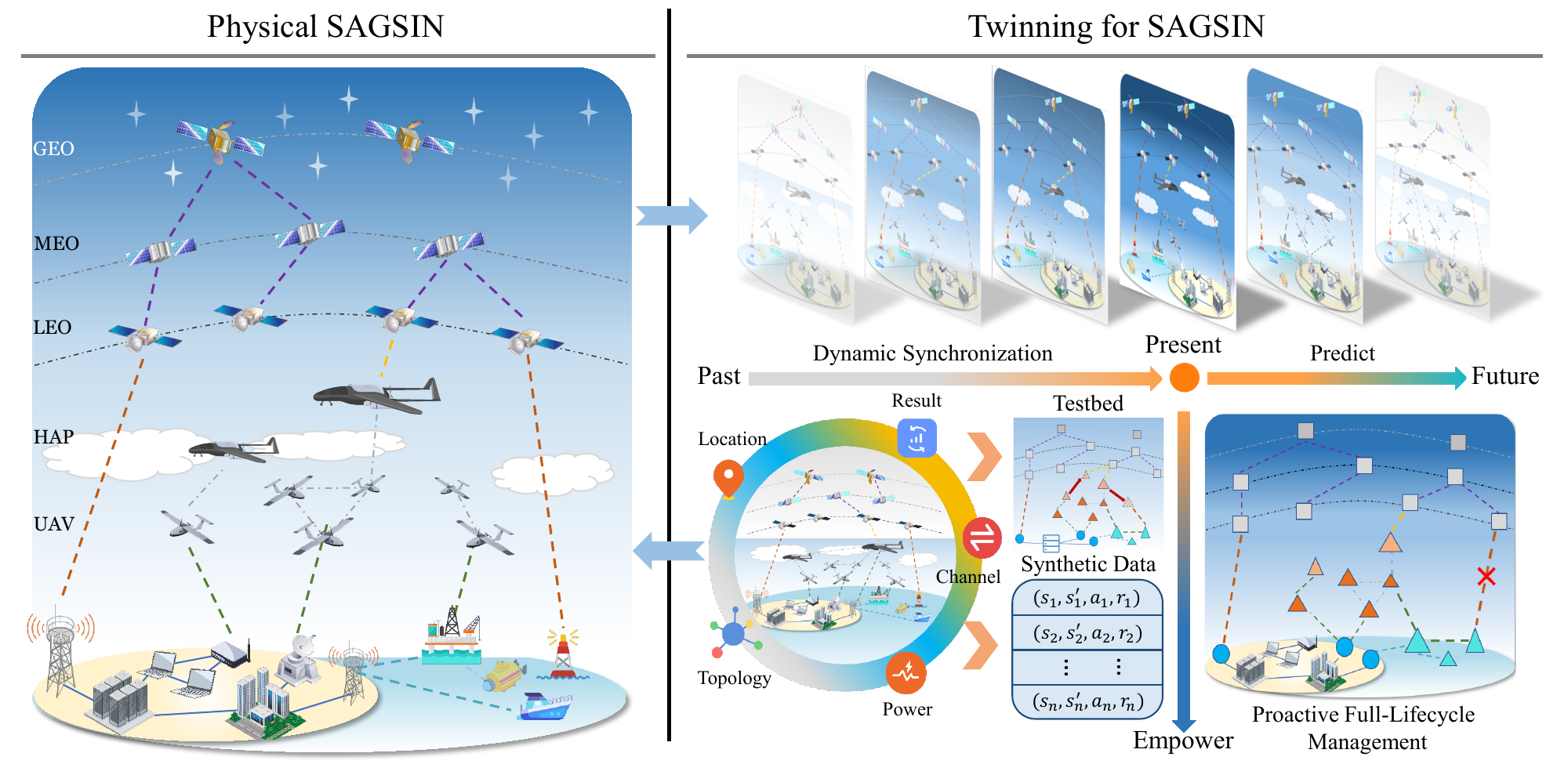} 
    \caption{The left side shows the physical SAGSIN, and the right side presents the key capabilities of a DT operating within SAGSIN. The DT maintains a continuously synchronized virtual environment through constant data interaction with the physical environment. Based on this virtual environment, the DT can predict changes in the SAGSIN, enabling proactive adjustments and the full-lifecycle management of the network \cite{chen2024survey}. Additionally, it can serve as a foundation for intelligent algorithms, providing virtual training environments or virtual samples at a low cost \cite{al2024digital, li2024automatic}.}
    \label{DT_faces_Twinning}
\end{figure*}

In recent years, routine human operations have expanded beyond terrestrial boundaries, extending from deep-sea exploration and low-altitude drones to ongoing activities in outer space. This expansion creates an imperative for highly reliable, low-latency connectivity. However, traditional terrestrial networks are primarily designed for densely populated areas and remain unavailable across approximately 70\% of the Earth's surface, failing to meet these large-scale, cross-domain communication requirements. In response, the space-air-ground-sea integrated network (SAGSIN) has emerged as a key architectural solution in 6G. As illustrated in Fig.~\ref{DT_faces_Twinning}, SAGSIN integrates  satellite networks, high-altitude platforms (HAPs), uncrewed aerial vehicles (UAVs), terrestrial networks and maritime networks into a unified framework, providing seamless global communication coverage.

Driven by market demand and ongoing standardization, SAGSIN is being rapidly deployed. Industry data indicates that the global market for low Earth orbit (LEO) satellite communications is growing at an average annual rate of over 20\%, with multiple mega-constellation projects entering intensive deployment phases \cite{Fortune2024MegaConstellations}. Meanwhile, international standards organizations such as the International Telecommunication Union (ITU) and 3GPP have formally initiated standardization work on satellite-terrestrial integration \cite{3GPP_TR36_763,3GPP_TR38_882}. In the foreseeable future, the scale of equipment and service diversity in SAGSIN will grow exponentially. Traditional network management solutions reliant on centralized data centers, human expertise, and static configurations will struggle to cope with sharply escalating complexity and mission volume. Consequently, \textbf{how to achieve efficient operation, maintenance, and orchestration within such a highly dynamic and heterogeneous architecture remains a fundamental open problem} ~\cite{wang2023efficient}. 
	
Digital Twin (DT) research aims to manage complex engineering systems by maintaining real-time virtual counterparts of physical assets, offering a viable solution to the aforementioned open problem. With a long history and extensive literature, DT has evolved through multiple conceptual iterations driven by diverse application requirements. This survey organizes its development into four overlapping stages:
First, entity-centric formulations tied a virtual counterpart to the identity and lifecycle of a physical asset, with early aerospace visions emphasizing highly detailed vehicle representations \cite{grieves2014digital,glaessgen2012digital}. Second, connected-twin studies made automated physical--digital data exchange, bidirectionality, synchronization, and lifecycle coverage central classification properties \cite{kritzinger2018digital,jones2020characterising,vanderhorn2021digital}. Third, deployment-oriented research introduced adaptive fidelity, surrogate modeling, edge execution, and fit-for-purpose scope to balance utility against cost \cite{rasheed2020digital,kober2022digital}. Fourth, recent semantic, cognitive, goal-aware, and formally verifiable twins have expanded the design space from representation accuracy to interoperability, task relevance, and justified behavior \cite{karabulut2024ontologies,li2023toward,alhazmi2026formal}. These stages are overlapping, not sequential: a connected twin may also be adaptive, and a semantic twin may also be formally verified.

Conceptually, DT holds significant potential for empowering SAGSIN, as illustrated in Fig.~\ref{DT_faces_Twinning}, and has thus attracted considerable attention. SAGSIN is intrinsically cross-domain: one service may traverse satellites, aerial platforms, terrestrial infrastructure, and maritime assets with different semantics, control cycles, links, and resource budgets. A maritime emergency, for example, can require a satellite relay, a UAV, and a terrestrial controller to interpret the same event despite intermittent and asynchronous observations. In practice, the large physical span between nodes, constrained communication links, sparse computational resources, and highly dynamic topologies all limit the applicability of conventional DT. The four developmental stages described above offer complementary insights into DT evolution; however, their individual scopes mean that these coupled challenges are rarely addressed in a unified manner. Existing DT research for SAGSIN remains largely confined to prospective architectural designs or small-scale simulations \cite{10107755,10979998,10669844}. This consequently exposes a systemic orchestration gap among these constituent components.  Isolated DT-centric innovations are therefore insufficient to meet the demands of future SAGSIN, and a systematic twinning approach tailored to such scenarios remains to be articulated.

To bridge the gap between conceptual vision and practical deployment, we leverage current DT trends and complementary technologies to design a twinning framework that meets SAGSIN requirements. We first analyze the structural mismatches between DT and SAGSIN, then propose moving beyond ``high fidelity'' and isolated technical innovations to explore two alternative developmental trends specifically suited for SAGSIN: \textbf{semantic} and \textbf{goal-oriented}. The semantic approach is premised on a unified knowledge base, moving beyond raw bits to provide a consistent representation for all system components; this facilitates data compression and sophisticated reasoning by leveraging intrinsic data relationships. Complementarily, the goal-oriented approach determines the modeling scope and precision directly from the task objective, thereby reducing resource demands. Together, these approaches form a synergistic framework that enables SAGSIN to achieve superior goal-oriented outcomes with greater data efficiency.

To distinguish such systems from existing works that extend conventional DT concepts in isolation,  we coin the term \textbf{goal-oriented semantic twin (GOST)} for use in the SAGSIN context, reflecting its emphasis on goal-directed reasoning and semantic relationships. GOST is defined by six coupled \emph{design components}: mission goal, semantic scope, resource envelope, physical--digital interaction policy, acceptance evidence, and lifecycle governance. They specify the required outcome, relevant cross-domain semantics, resource constraints, synchronization policy, action justification, and instance accountability. Semantic communication, ontologies, lightweight modeling, and formal methods implement these decisions. This survey offers a comprehensive treatment of the motivation, definition, boundaries, and interrelationships of this concept, together with detailed descriptions of the key enabling technologies and deployment processes.

\subsection{Related Surveys \& Magazines}

\revRonetwo{The closest surveys fall into four groups: SAG(S)IN architecture, security, and resource management; broad DT characterization, modeling, and enabling technologies; semantic-enabled DT research; and DT-enabled SAGIN work.} The six-component GOST framework (mission goal, semantic scope, resource envelope, physical--digital interaction policy, acceptance evidence, and lifecycle governance) is mapped onto seven comparison columns; the interaction-policy component is reflected jointly in the Resource and Lifecycle columns because interaction granularity and synchronization rate are determined by both the resource envelope and the governance policy. Table \ref{tab:survey_comparison} compares representative surveys directly rather than inferring novelty from a self-constructed chronology. The comparison asks whether each work jointly covers SAGSIN constraints, semantic interoperability, goal-derived scope, resource-adaptive deployment, lifecycle governance, multidimensional evaluation, and an end-to-end case. Existing surveys provide strong coverage within individual groups, but none treats these seven dimensions as one composition rule for SAGSIN.


\subsubsection{Architecture, Composition, and Characteristics of SAG(S)IN}
In the macro-vision and architectural design of 6G, \cite{Saad6Gvison} approaches 6G from its driving applications, technological trends, and enabling technologies, establishing SAGIN as one of the key architectures and emphasizing its multi-dimensional performance requirements for high reliability, low latency, and high data rates. Building on this foundation, \cite{liu2018space} examines in greater depth the core network challenges of SAGIN, systematically reviewing issues such as heterogeneity, dynamic topology, and high latency in network design, resource allocation, and protocol optimization, while also exploring traditional network solutions like SDN/NFV.

As the architecture becomes more defined, research has begun to shift towards specific enabling technologies and key sub-domains. In the area of intelligent O\&M, \cite{jia2024service} explores an edge intelligence (EI)-driven SAGIN architecture, emphasizing the use of deep reinforcement learning (DRL) for resource management and computation offloading. In the domain of wireless communication, \cite{cao2025survey} conducts an in-depth study focused on near-space communications (NS-COM) within SAGSIN, with a detailed analysis of its channel characteristics and networking optimization.

Although these studies do not directly address twinning technology, their reliance on high-frequency environmental interaction and massive data indirectly corroborates the need for twinning technology as a virtual training environment.

\subsubsection{The Development of Twinning Technology} The authors in \cite{jones2020characterising} offer an in-depth analysis differentiating DT from modeling and simulation (M\&S) based on bidirectional physical interaction. \cite{mihai2022digital} surveys common DT concepts to define their scope and functional boundaries. 

With this conceptual foundation in place, the core emphasis of DT research has shifted toward the discussion of its practical deployment. Mashaly \cite{mashaly2021connecting} notes that the core value of a DT lies in the real-time bidirectional data synchronization between the physical entity and the virtual model, while network latency can directly undermine its applicability in time-sensitive environments.

\begin{table*}[t]
\centering

\caption{Direct Comparison with Representative Surveys}
\label{tab:survey_comparison}
\scriptsize
\setlength{\tabcolsep}{3.3pt}
\renewcommand{\arraystretch}{1.12}
\begin{tabularx}{\textwidth}{>{\RaggedRight}p{2.5cm}>{\RaggedRight}p{2.15cm}ccccccc>{\RaggedRight}X}
\toprule
\textbf{Work} & \textbf{Primary scope} & \textbf{SAGSIN} & \textbf{Sem.} & \textbf{Goal} & \textbf{Res.} & \textbf{Life.} & \textbf{Eval.} & \textbf{Case} & \textbf{Main boundary relative to this survey} \\
\midrule
Liu et al. \cite{liu2018space} & SAGIN survey & Y & -- & -- & P & -- & P & -- & Architecture, routing, and resource management without twinning. \\
Guo et al. \cite{Guo2022asurveyon} & SAGSIN security & Y & -- & -- & P & P & P & -- & Network threats and defenses rather than semantic twin lifecycle risks. \\
Chen et al. \cite{chen2024survey} & JCC-SAGIN resources & Y & -- & P & Y & -- & Y & -- & Communication--computing resource management without a twin composition rule. \\
Jones et al. \cite{jones2020characterising} & DT characterization & -- & -- & -- & P & Y & P & -- & Defines DT properties and lifecycle concepts across domains. \\
Rasheed et al. \cite{rasheed2020digital} & DT modeling & -- & -- & P & Y & P & Y & P & Analyzes fidelity--cost tradeoffs but not SAGSIN semantics. \\
Mihai et al. \cite{mihai2022digital} & DT technologies & -- & P & P & Y & Y & Y & P & Broad DT taxonomy and trust challenges without SAGSIN joint design. \\
Karabulut et al. \cite{karabulut2024ontologies} & Ontologies in DT & -- & Y & P & -- & P & P & -- & Deep ontology review without goal/resource co-instantiation. \\
Cui et al. \cite{cui2022sagin} & DT-enabled SAGIN & Y & -- & P & Y & P & P & P & Adapts SAGIN infrastructure to DT traffic and synchronization. \\
\textbf{This survey} & \textbf{GOST for SAGSIN} & \textbf{Y} & \textbf{Y} & \textbf{Y} & \textbf{Y} & \textbf{Y} & \textbf{Y} & \textbf{Y} & \textbf{Integrated framework linking goal, semantics, resources, interaction, evidence, and lifecycle.} \\
\bottomrule
\end{tabularx}
\vspace{1mm}
\begin{minipage}{0.98\textwidth}\scriptsize
Y: systematic coverage; P: partial or supporting coverage; --: outside the stated primary scope. Sem.: semantic interoperability; Goal: goal-derived scope; Res.: resource-adaptive deployment; Life.: lifecycle/governance; Eval.: evaluation framework.
\end{minipage}
\end{table*}

To address the challenges posed by timeliness, heterogeneity, and collaboration, the research frontier has begun to turn toward semantic technologies and goal-oriented approaches. To enable efficient collaboration among DTs, ontologies play a crucial role in this process. \cite{karabulut2024ontologies} utilizes a knowledge graph (KG) to manage metadata and defines the modeling scope through backward derivation of task objectives; \cite{listl2024knowledge} employs KG to seamlessly integrate heterogeneous data sources in the field of industrial automation. Li \textit{et al.} \cite{li2023toward} propose the specific concept of semantic-enhanced digital twin (SDT), which combines task objectives to extract semantic knowledge---such as geometric and behavioral features---to achieve a holistic understanding of interactive behaviors, thereby optimizing timeliness and computational costs.

Existing research outlines a development path for DT that evolves from M\&S toward real-time synchronization as its core value, yet high fidelity and physical--virtual interaction remain challenging \cite{jones2020characterising}. While semantic technologies and goal-oriented modeling have shown considerable promise for real-time twinning, prior studies tend to treat these elements as disjoint enablers. Consequently, how to jointly organize them into a cohesive architecture for multi-domain SAGSIN operation has yet to be systematically investigated.

\subsubsection{Twinning Technologies in SAG(S)IN} There is significant interest in integrating twinning technology with SAGSIN. Cui \textit{et al.} \cite{cui2022sagin}, from a network infrastructure perspective, point out that the high latency and hardware rigidity of 5G NTN cannot meet the bidirectional synchronization demands of DT. To address this, they propose a novel SAGIN architecture that reduces latency and enhances flexibility by distributively deploying core network functions (e.g., migrating them to satellites). Chen \textit{et al.} \cite{chen2024survey} approach the problem from an application-enablement perspective, emphasizing the potential of DT combined with AI and mobile edge computing (MEC) to address time-varying topologies and optimize resource management in JCC-SAGIN.

\begin{figure*}[htbp]
    \centering
    \includegraphics[width=0.95\textwidth]{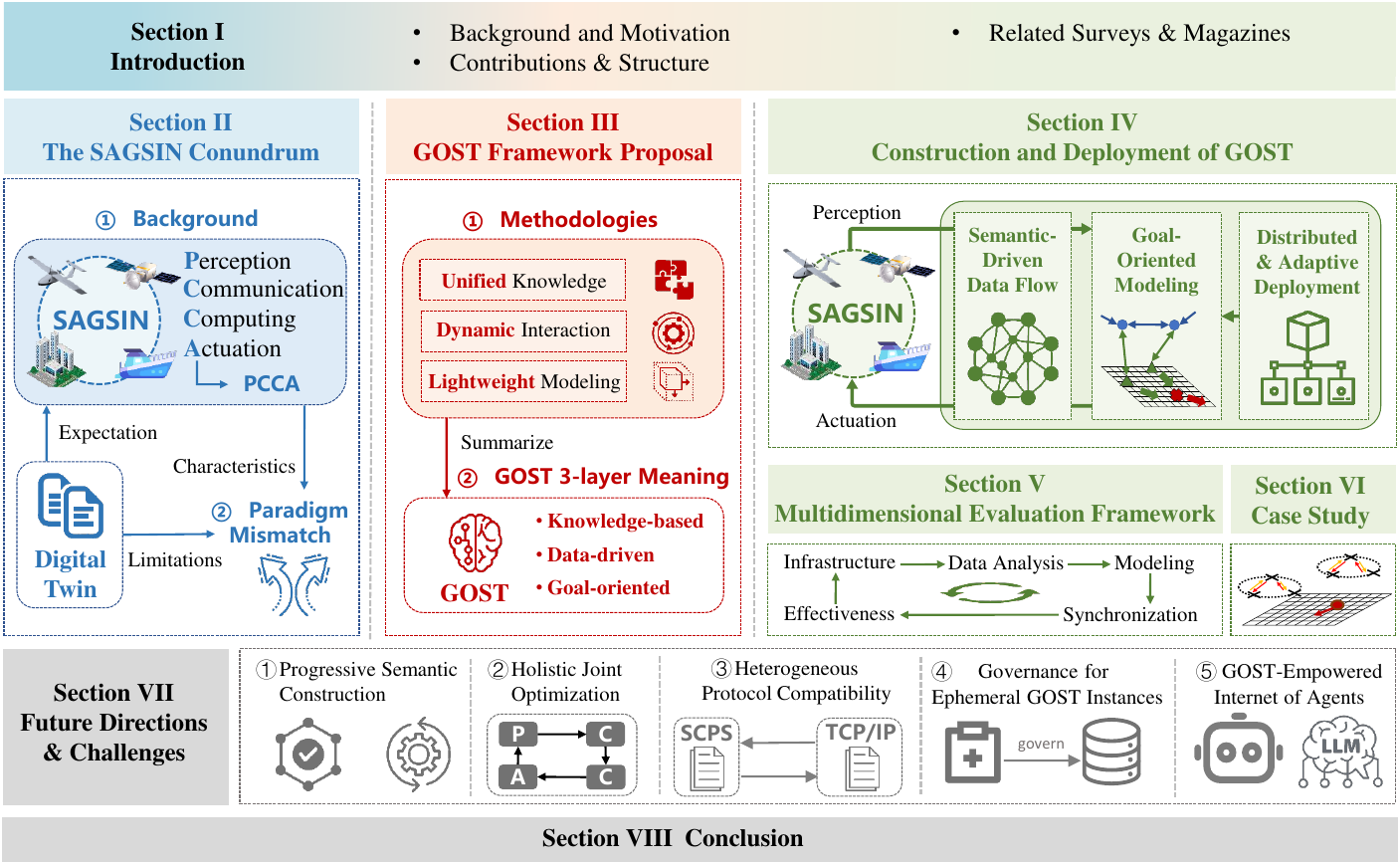} 
    \caption{Survey Structure. Sections II and III summarize the conditional mismatch between DT and SAGSIN and, based on a broad exploration of new paradigms, propose the GOST framework. Sections IV through VI then serve as a tutorial for constructing GOST. Finally, Section VII discusses future research directions for GOST, and Section VIII concludes the paper.}
    \label{figlabel1}
\end{figure*}

The gap is therefore not that prior surveys overlooked every ingredient of GOST. Rather, SAG(S)IN surveys generally stop at architecture, security, or resource optimization; DT surveys span enabling technologies but are not organized around SAGSIN's intermittent and asymmetric PCCA loop; and semantic twin surveys do not make runtime goal, resource, assurance, and retirement decisions co-dependent. This survey addresses that precise intersection and retains the mature DT literature as its technical foundation.

\subsection{Contributions \& Structure}
 

We characterize SAGSIN as a continuous \textit{perception--communication--computing--actuation} (PCCA) loop \cite{meng2024semantics} and use it to identify where entity-wide, continuously synchronized twin implementations become costly or stale. We then characterize GOST as an integrated framework with six interdependent components---mission goal, semantic scope, resource envelope, physical--digital interaction policy, acceptance evidence, and lifecycle governance---and review the technologies and evaluation methods needed to implement their joint design logic.

With a focus on GOST within the context of SAGSIN, the contributions of this paper are summarized as follows:

\begin{itemize}

\item Through the PCCA-loop lens, we identify the fundamental challenges confronting twinning systems in SAGSIN, systematically analyze the root causes of structural mismatch between SAGSIN requirements and conventional DT paradigms, and distill the defining characteristics that GOST must embody.

\item We define GOST as an integrated six-component framework and distinguish it from adaptive/lightweight, semantic/cognitive, goal-oriented, and formally trustworthy DT families.

\item We provide a construction tutorial that covers progressive semantic bootstrapping, task-aware communication, multi-goal coordination, adaptive deployment, semantic robustness, and lifecycle provenance.

\item We organize evaluation across five PCCA-aligned dimensions and a cross-cutting trustworthiness plane spanning security, conflict handling, and auditability.

\item We present a remote multi-UAV tracking case that connects semantic synchronization and prioritized mission objectives to measurable state freshness and task utility.

\end{itemize}

The rest of this paper is structured as follows: Section II summarizes the characteristics of SAGSIN via the PCCA loop and provides an in-depth analysis of the mismatch between the conventional DT and SAGSIN. Section III then reviews the explorations of new twinning paradigms and ultimately proposes the GOST framework. Section IV centers on the proposed GOST, detailing its construction and deployment approaches in the context of SAGSIN. To address the relative lack of research on the evaluation of twin models, Section V focuses on summarizing evaluation methods and metrics for GOST. Section VI presents a case study to validate the proposed framework. Section VII discusses ongoing challenges and future research directions pertaining to GOST. Section VIII concludes the paper. An overview of the organization and structure of this paper is illustrated in Fig. \ref{figlabel1}.

\section{The SAGSIN Conundrum: Architectural Complexity and the Mismatch with Digital Twin Paradigm} \label{perfect_twin}

This section uses the PCCA loop to characterize SAGSIN and then identifies the conditions under which entity-wide scope, broad synchronization, and fixed deployment create a mismatch. These are limitations of a design configuration, not of every system described as a DT.



\subsection{Characteristics of SAGSIN within the PCCA Loop}

SAGSIN is a foundational 6G architecture envisioned as a unified, vertically heterogeneous ``network of networks'' designed for ubiquitous global connectivity. Its core design principle is the deep integration of disparate space-based assets (multi-layered satellite constellations), air-based platforms (HAPs/UAVs), high-capacity terrestrial cellular networks, and sea-based maritime nodes into a single, cohesive system \cite{chen2025space}.

\begin{figure*}[htbp]
    \centering
    \includegraphics[width=0.95\textwidth]{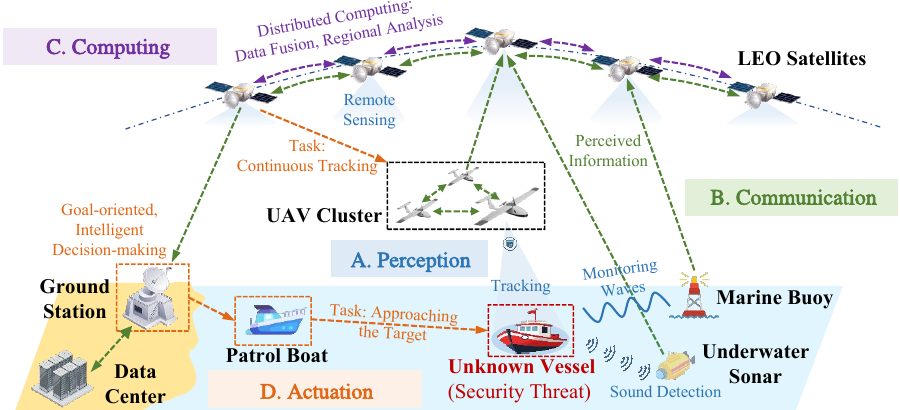} 
    \caption{Using UAV tracking as an example: after the UAV perceives a target, the data is transmitted to the satellite constellation. Following distributed computing within the satellite constellation and processing by the ground station, control commands are sent to the ship and the UAV. The task under SAGSIN is completed through the continuous execution of the PCCA loop \cite{meng2024semantics}.}
    \label{SAGSIN_PCCA_pic}
\end{figure*}

As illustrated in Fig. \ref{SAGSIN_PCCA_pic}, SAGSIN intelligence is organized as a continuous PCCA loop. Perception converts sparse observations from heterogeneous satellites, platforms, vehicles, and maritime sensors into data streams. Communication must carry these streams across rapidly changing topologies, Doppler-impaired links, and propagation delays ranging from terrestrial to satellite scales \cite{wang2021hybrid,lyu2024dynamic,You2020Massive,cui2022sagin,lin2021doppler,meng2024semantics,zhao2025satpipe}. Computing is consequently distributed toward UAVs, HAPs, and satellites, where SWaP limits make placement and offloading depend jointly on latency, link stability, energy, and available computation \cite{cui2022sagin,abrar2021energy,mcenroe2022survey}. Actuation closes the loop through scheduling, trajectory, and interference-control decisions \cite{qiu2022scheduling,qiu2024timely,dong2023drl,wang2021incorporating,liu2022energy}. Feedback from these decisions returns to perception, while heterogeneity in sensors, protocols, formats, and resources affects every transition and raises the cost of maintaining a coherent global state \cite{elmahallawy2024communication}.


\subsection{A Conditional Mismatch: Fidelity-Led Twinning in the SAGSIN Context} \label{dt_conflict}

DT has achieved substantial results in manufacturing and aerospace \cite{li2021digital}, and adaptive-fidelity, edge, semantic, and goal-aware variants already relax individual limitations of early formulations. Nevertheless, a persistent system-level mismatch remains widespread in SAGSIN deployment: twin operation must jointly reconcile dense observation demands, broad and frequent synchronization, entity-wide model scope, and heterogeneous resource availability. Across space, air, ground, and sea segments, these requirements repeatedly encounter sparse sensing, intermittent links, asynchronous state, rapidly changing topology, and asymmetric computing resources. As SAGSIN services expand across domains and become more latency-sensitive, this unresolved tension is likely to become a major bottleneck to scalable twinning deployment.

\subsubsection{Scope Conflict: Sample Acquisition Cost vs. Entity-Wide Replication} A fidelity-led, entity-wide twin requires broad and diverse observations. For example, a detailed spacecraft twin may draw on scan and engineering data from a large fraction of the physical asset \cite{li2021digital}. Few DTs require this full scope in principle; the practical problem is that such scope cannot be treated as fixed before the SAGSIN task and resource envelope are known.

SAGSIN operates on a global scale, with devices sparsely distributed across its coverage area. Perception units in the PCCA loop cannot monitor every location, and this limitation is exacerbated by fragile communication channels, which degrade the quality and quantity of data reaching the DT. Moreover, the extreme heterogeneity of SAGSIN further strains the PCCA loop: diverse node types increase the DT modeling workload, while divergent data formats and communication protocols raise the cost of sample acquisition \cite{cui2022sagin}.

In this context, the limitation of DT's ``mirror replication'' paradigm surfaces at the very first step of the PCCA loop: conventional DT assumes that data is abundantly available, whereas the physical state of SAGSIN is sparse and costly to acquire.

\subsubsection{Interaction Conflict: Broad Synchronization vs. High Dynamics and Large Scale} Physical--digital interaction remains a defining DT property, but neither all variables nor all tasks require the same update rate \cite{mihai2022digital,han2022dynamic}. A SAGSIN implementation that attempts broad, lossless, and uniformly frequent synchronization nevertheless requires persistent high-bandwidth links and can deliver stale state to the actuation stage. GOST therefore treats update scope, semantic importance, and tolerated age as interdependent design choices within the framework.

Firstly, the limited network bandwidth, storage, and processing resources in SAGSIN make it difficult to handle large-scale information exchange between data terminals and their physical counterparts \cite{van2022edge}. Due to factors such as Doppler shift, inter-node synchronization, and low bandwidth utilization, transmission delays are prolonged and subject to pronounced uncertainty, while success rates remain low.

Secondly, the highly dynamic nature of SAGSIN causes the value of information to degrade rapidly over time. For example, the SAGSIN topology can be reconfigured on a second-level timescale \cite{wang2021hybrid,lyu2024dynamic}, and the coherence time of the channel is extremely short. In extreme cases, the topology information and channel state information (CSI) within the PCCA loop may become outdated before being used, failing to reflect the physical environment.

Furthermore, SAGSIN exhibits ``heterogeneous and asynchronous delays.'' Due to the large scale and heterogeneous characteristics of SAGSIN, the state data received by the digital twin at the same moment correspond to different time points of the physical entities \cite{10879302}, significantly increasing the complexity of digital twin management.

\subsubsection{Deployment Conflict: Model Demand vs. Resource Asymmetry} Detailed simulation and large historical stores are commonly placed on cloud or high-performance edge servers. Rasheed \textit{et al.} identify computational efficiency as a central DT enabler \cite{rasheed2020digital}. In SAGSIN, however, the nodes that must close the fastest control loops may be satellites or UAVs with the least energy, memory, and compute. The relevant design question is thus not whether DTs can be lightweight, but whether model scope and placement are recomputed with the task and current resource envelope.


At the computing stage, many detailed DT deployments place the digital model on infrastructure more capable than the physical endpoint. In SAGSIN, the endpoint may also host the latency-critical model despite strict resource limits. This asymmetry motivates joint selection of scope, compression, placement, and fallback rather than a claim of fundamental DT incompatibility.


These conflicts do not establish the inapplicability of DT as a field. They identify a missing composition rule for SAGSIN. Adaptive and lightweight DTs mitigate model cost \cite{zhang2022improved,vanderhorn2021digital,kim2022data,cao2021simulation}; semantic twins improve interoperability; goal-oriented models select task-relevant scope; and formal methods strengthen assurance. GOST requires these choices to be made together, because reducing fidelity alone does not determine which state to acquire, what to transmit, where to execute, when to retire an instance, or what evidence makes its action acceptable.

\subsection{Lessons Learned}
The PCCA analysis characterizes SAGSIN by large scale, high dynamics, extreme heterogeneity, and computational scarcity \cite{cui2022sagin,lyu2024dynamic,wang2021hybrid}. It does not reject DT; it shows why entity-wide scope, uniform synchronization, and fixed placement should not be default choices in a data-scarce, high-latency, asynchronous system \cite{rasheed2020digital,gao2024cost}.

Therefore, the future frontier of intelligent SAGSIN management should no longer be confined to achieving finer-grained physical simulations. Instead, research should focus on developing foundational theories and methods for \textit{goal-oriented design}, \textit{semantic modeling}, and \textit{resource-adaptive allocation}. 



\section{From Digital Replicas to Semantic Intelligence and Goal-oriented Principle} \label{section iii}

This section first reviews the explorations of new twinning frameworks by summarizing their research trends and advanced concepts. Subsequently, we propose the GOST concept through a three-layer framework, detailing the roles of semantic technology and goal-oriented principles within GOST and clarifying how it addresses the limitations of SAGSIN.

    \begin{table*}[htbp]
	\centering
	\caption{Summary of New Twinning Paradigm Explorations}
	\label{tab:new_twinning_paradigms_en_vcenter_4col} 
	\renewcommand{\arraystretch}{1} 
	
	\begin{tabularx}{\textwidth}{
			>{\RaggedRight}p{2.7cm} 
			>{\RaggedRight}p{5cm} 
			>{\RaggedRight}X 
			>{\RaggedRight}p{2cm}
		}
		\toprule
		\textbf{Paradigm} & \textbf{Objective} & \textbf{Main Methods / Approaches} & \textbf{Related Ref.} \\
		\midrule
		
		\multirow{2}{=}{\RaggedRight \textbf{Unification of Modeling Languages}}
		& To address heterogeneity and establish unified syntactic frameworks for entities and networks.
		& \textbf{Establishing unified syntactic/ontological frameworks} to define entities, attributes, and relationships.
		& \cite{zhao2022digital}, \cite{10103508}, \cite{kulikov2021ontology} \\
		\cmidrule(lr){2-4} 
		& To achieve semantic interoperability and autonomous interpretation of heterogeneous data.
		& \textbf{Using AI to achieve semantic interoperability}; moving from rule-based syntax to knowledge-based semantic understanding.
		& \cite{rico2023context}, \cite{11078840}, \cite{yu2023digital} \\
		
		\midrule
		
		\multirow{3}{=}{\RaggedRight \textbf{Discovery of Data Associations}}
		& To address data scarcity, high dynamics, and latency via compression in constrained environments.
		& \textbf{Compressing raw data based on task relevance} (semantic compression) by extracting task-relevant features to reduce transmission load.
		& \cite{letaief2021edge} \\
		\cmidrule(lr){2-4}
		& To extract spatiotemporal patterns from minimal data.
		& \textbf{Extracting spatiotemporal patterns} from data using hybrid neural networks to understand underlying relations without complex preprocessing.
		& \cite{zhu2021knowledge} \\
		\cmidrule(lr){2-4}
		& To reduce dependency on large-scale real data by generating virtual samples.
		& \textbf{Using GAI to create synthetic data} to compensate for data scarcity and improve ML model training.
		& \cite{10409284}, \cite{huang2024digital}, \cite{chai2024generative} \\
		
		\midrule
		
		\multirow{2}{=}{\RaggedRight \textbf{Design for Targeted Objectives}}
		& To shift from high-fidelity to a lightweight, ``fit-for-purpose'' principle guided by task demands.
		& \textbf{Adopting a ``fit-for-purpose'' (goal-oriented) approach} to model only task-relevant subsets, reducing complexity and resources.
		& \cite{zhao2022digital}, \cite{raja2024towards}, \cite{raja2023using} \\
		\cmidrule(lr){2-4}
		& To enable lightweight, task-specific twin allocation and automated network strategies.
		& \textbf{Using KG to implement intent-driven networking}, automatically matching task goals (intent KG) with network states (state KG) to generate strategies.
		& \cite{letaief2021edge}, \cite{chang2022kid} \\
		
		\bottomrule
		\end{tabularx}
	\end{table*}

\subsection{Exploration of New Twinning Frameworks}

	Table \ref{tab:new_twinning_paradigms_en_vcenter_4col} organizes the literature into three complementary directions: unified representations for heterogeneous entities, data association and efficient transmission under scarcity, and task-driven designs that allocate modeling effort where it matters.

	\subsubsection{Unification of Modeling Languages}
	Unified modeling studies use ontologies to specify entities, attributes, and relationships \cite{zhao2022digital}; organize mobile-network twinning into acquisition, modeling, adaptation, deployment, provisioning, interconnection, and feedback \cite{10103508}; and describe heterogeneous equipment and dynamic transmission through telecom monitoring ontologies \cite{kulikov2021ontology}.

Building on these foundations, related research has progressed toward semantic interoperability for autonomous interpretation of heterogeneous data within SAGSIN. A key characteristic of these efforts is the use of AI algorithms to move from rule-based syntactic approaches to knowledge-based semantic understanding. \cite{rico2023context} shares a workflow similar to \cite{zhao2022digital} but places greater emphasis on cross-domain and cross-task scenarios by incorporating richer contextual information into the ontology. The authors in \cite{11078840} propose the concept of a semantic twin network, leveraging large language models (LLMs) to extract semantic information from multi-source heterogeneous data, thereby populating static network features such as device information and topology structure. This enables SAGSIN to be described and reasoned about through text. Yu \textit{et al.} in \cite{yu2023digital} apply twinning technology to detect anomalies in 6G edge networks or links. They abstract the physical network and services as an undirected graph and a directed acyclic graph (DAG), respectively. The key network components---such as links, queues, and flows---are represented as nodes, and their interactions are modeled through multiple relation types, including link–flow, queue–flow, flow–queue, and queue–link connections.

\subsubsection{Discovery of Data Associations}
Under sparse data and constrained transmission, studies combine task-aware compression, semantic communication, and distributed learning. Letaief \textit{et al.} define task objectives and metrics, extract relevant features with DNNs, and transmit them through JSCC, while Zhu \textit{et al.} recover spatiotemporal relations from sparse samples with an LSTM--CNN network relation extractor \cite{letaief2021edge,zhu2021knowledge}. Generative approaches use GAN, GAIL, diffusion models, and reinforcement learning to expand network scenarios and predict future states \cite{10409284,huang2024digital}; GMNDT similarly combines diffusion sampling with antenna-control learning and reports 6.91\% higher coverage and 29.8\% higher throughput than conventional methods \cite{chai2024generative}.


	
	\subsubsection{Design for Targeted Objectives}
	Goal-oriented studies replace universal replication with fit-for-purpose scope and precision. Task metrics can guide transmission design \cite{letaief2021edge}; device twins can be distinguished from lightweight service twins for edge allocation \cite{9491087}; and robot or manufacturing twins can restrict modeling to actions and stages relevant to the task \cite{raja2024towards,raja2023using}.

    \begin{figure}[t]
    \centering
    \includegraphics[width=0.48\textwidth]{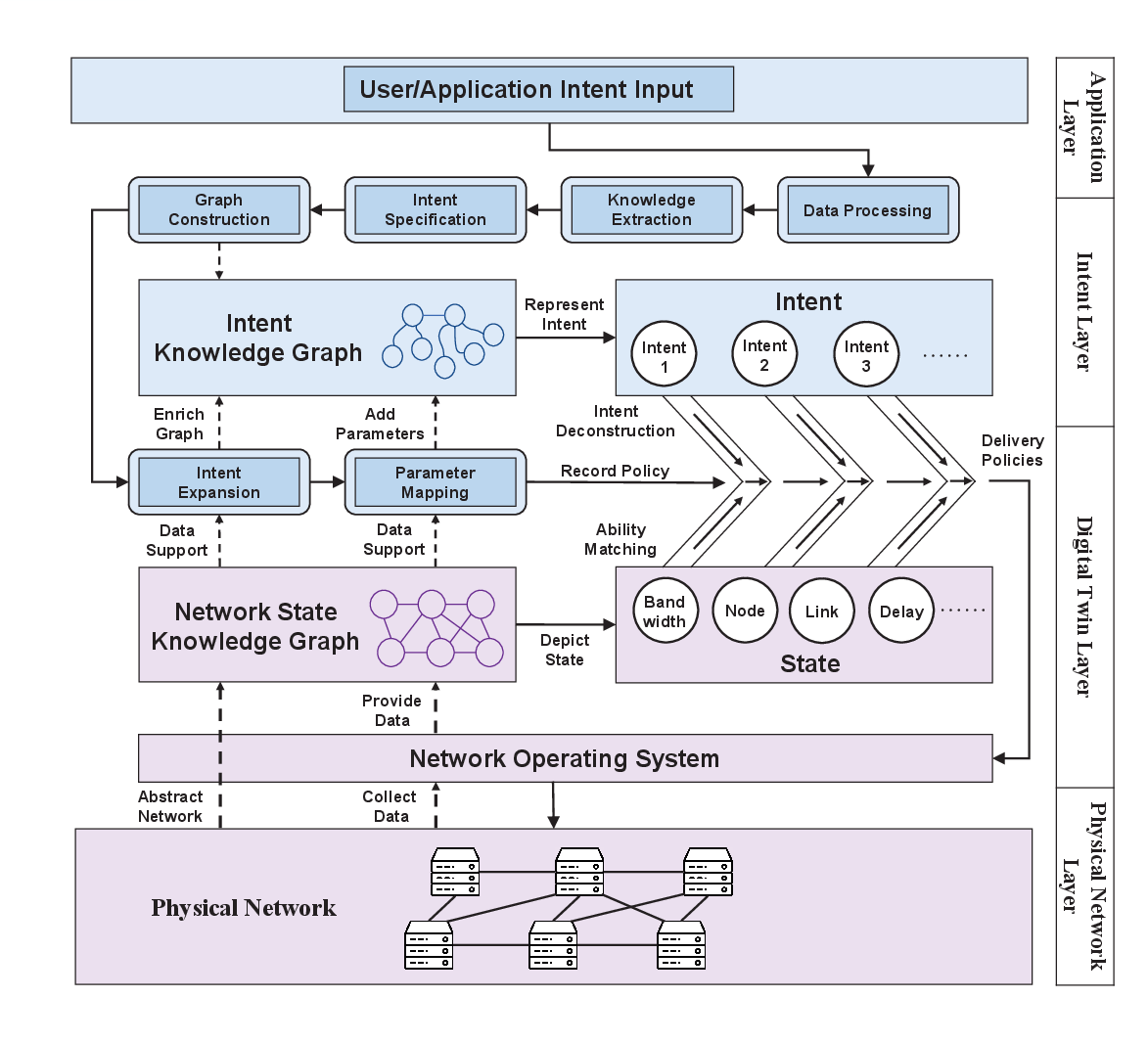} 
    \caption{  An illustration of a novel SAGSIN twinning paradigm leveraging KG to address heterogeneity and goal-oriented principles to enhance autonomous intelligence \cite{letaief2021edge,zhao2022digital}.}
    \label{kid_semantic_twin}
\end{figure}

Goal-oriented modeling also links intent to network operations. Backpropagating task semantics can produce directed task chains \cite{zhao2022digital}, while KID matches an intent KG with a network-state KG to derive strategies from requirements, topology, latency, and bandwidth \cite{chang2022kid}.

As summarized in Fig. \ref{kid_semantic_twin}, the literature follows two complementary directions: bottom-up semantic mapping of nodes and relations for unified PCCA representation, and top-down decomposition of objectives into sub-goals and controllable variables.

\subsection{Toward Goal-Oriented Semantic Twins: The Three-Layer Framework and Definition}

From the above explorations into new paradigms, it is evident that semantic technologies and goal-oriented thinking represent major innovation trends. However, relevant research still predominantly uses ``digital twin'' as its keyword, and much emerging work remains constrained by traditional DT definitions, which limits progress to incremental innovation. Therefore, this paper introduces the concept of GOST, where the focus shifts from the isomorphic replication of physical systems to constructing task-relevant semantic representations.

\begin{figure}[t]
    \centering
    \includegraphics[width=0.45\textwidth]{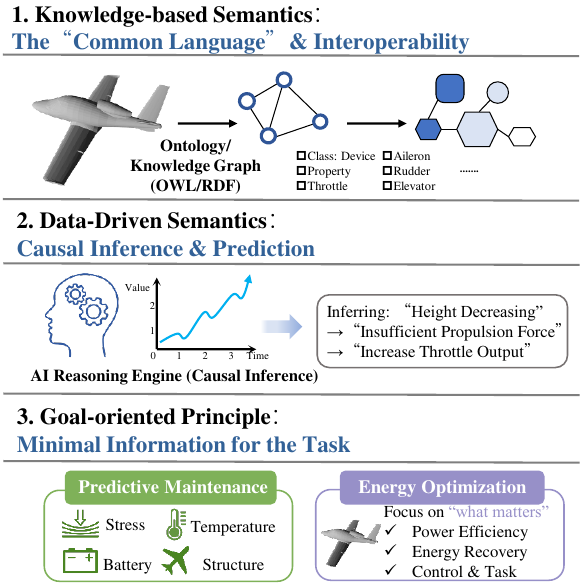} 
    \caption{Top-down functions of the three GOST layers: providing unified twinning workflows for heterogeneous nodes, characterizing dynamic SAGSIN under data sparsity, and simplifying models while enabling autonomy.}
    \label{figlabel2}
\end{figure}

To gain a deeper understanding of GOST, we decompose it into multiple layers of meaning. As the example in Fig. \ref{figlabel2} illustrates, it forms a framework that spans data structures, intelligent reasoning, and task-specific objectives. These three layers can be configured in diverse combinations to construct GOST models with different granularities and intelligence levels.

\subsubsection{Knowledge-Based Semantics} This layer aims to create a unified, reusable, and machine-readable knowledge foundation for the entire SAGSIN. It constitutes the knowledge-based mapping and serves as the ``master blueprint'' for all subsequent GOST instances. Using semantic web technologies such as ontologies (e.g., OWL\footnote{https://www.w3.org/OWL/}) and knowledge graphs (e.g., RDF\footnote{https://www.w3.org/RDF/}), it provides a unified, formal, and machine-readable vocabulary and data model for all entities within the system---including devices, sensors, processes, and products---and their interrelationships \cite{rico2023context,karabulut2024ontologies,Men2025Inter}. For SAGSIN, a unified semantic repository analogous to those in \cite{zhao2022digital,10103508,kulikov2021ontology} is essential, serving to uniformly represent nodes across different organizations and their interconnections during GOST modeling. This layer involves mapping massive, heterogeneous real-time data streams---from satellites, UAVs, ground and sea facilities, etc.---to corresponding entities and attributes in the KG.

Although sufficient shared semantic knowledge is essential for interoperable deployment, a repository need not be complete before operation begins. In practical deployments, semantic construction commonly follows a progressive cold-start workflow. A small upper ontology can first fix only identifiers, units, spatial--temporal relations, and provenance fields. Local operators can retain domain ontologies and publish mappings rather than raw operational data, while federated aggregation can align candidate entities and relations into a global semantic view, following the privacy-preserving collaboration principle used in distributed DT learning \cite{lu2020communication}. When inconsistent mappings arise across operators, they can remain explicit candidates and be resolved through declared inter-operator validation rules rather than automatic majority voting. LLMs can propose classes, relations, and mappings from technical documents or telemetry schemas \cite{11078840}, but every proposal should remain provisional until shape constraints, physical rules, competency questions, and domain experts validate it. If the constraint-violation rate exceeds a deployment-specific threshold, the deployment can narrow the automated scope or fall back to manually curated templates. Finally, versioned incremental updates can add or retire concepts without rebuilding the repository, while dependency and consistency checks propagate only affected changes \cite{9813407,Qian_Wang}. Together, these stages support an evolving knowledge service instead of treating a monolithic repository as a one-time construction requirement.

	\subsubsection{ Data-Driven Semantics} The objective of this layer is to infuse autonomous intelligence into the static knowledge blueprint established by the knowledge-based layer. The primary approach involves mining the underlying associations within the data---including spatiotemporal features, causal relationships, and information significance---thereby addressing the challenges of data sparsity and constrained communication within the SAGSIN PCCA loop.
	
	Although SAGSIN exhibits highly dynamic characteristics, strong correlations exist between prior and subsequent states across both temporal and spatial scales; for instance, satellites, despite their high velocity, follow predictable trajectories. By leveraging dynamic models and algorithms such as LSTM to extract spatiotemporal features from various components, GOST can utilize data from key nodes to perform temporal state prediction and spatial state completion \cite{zhu2021knowledge}.

	Besides reducing communication pressure, exploiting correlations among sample data can also alleviate sample sparsity in SAGSIN. For example, causal inference algorithms capture state transition laws \cite{thomas2023causal}, enhancing modeling accuracy under sparse data conditions; physics-informed neural networks (PINNs) can infer unobserved states based on prior physical laws \cite{kim2022data}; and GAI methods, as discussed in \cite{huang2024digital,10409284,chai2024generative}, can generate virtual SAGSIN samples.
	

\begin{table*}[htbp]
\centering
\caption{Conceptual Boundary Between Modern DT Families and GOST}
\label{tab:dt_vs_st}
\small
\renewcommand{\arraystretch}{1.12}
\begin{tabularx}{\textwidth}{>{\RaggedRight}p{2.8cm}>{\RaggedRight}p{4.8cm}>{\RaggedRight}X}
\toprule
\textbf{Family} & \textbf{Primary design commitment} & \textbf{Boundary relative to the GOST framework} \\
\midrule
Entity-centric connected DT \cite{kritzinger2018digital,jones2020characterising,vanderhorn2021digital} & Maintains an identifiable physical counterpart, automated data exchange, and lifecycle synchronization. & May use fixed entity scope and fidelity-led evidence; goal, resources, semantic relevance, and retirement need not be co-determined. \\
Adaptive/lightweight DT \cite{rasheed2020digital,kober2022digital,Xiaokang_Zhou} & Trades fidelity, model size, or placement against computation and communication cost. & Resource adaptation is central, but semantic relevance need not govern acquisition, communication, acceptance, and instance termination together. \\
Semantic/cognitive DT \cite{karabulut2024ontologies,11078840,li2023toward} & Uses ontologies, KGs, or learned semantics for interoperability, reasoning, and compressed interaction. & Semantics may enrich a persistent entity-defined twin without making task scope and lifecycle conditional on runtime goals and resources. \\
Goal-oriented DT \cite{9491087,raja2024towards,raja2023using} & Derives model content or fidelity from requirements and intended use. & Goal modeling may stop at design time and need not control runtime semantic acquisition, synchronization, evidence, and retirement. \\
Formal/trustworthy DT \cite{mihai2022digital,alhazmi2026formal} & Specifies, verifies, or audits twin behavior and trust properties. & Formal assurance determines whether a chosen abstraction behaves correctly, but not whether the abstraction is task-relevant or resource-feasible. \\
\textbf{GOST} & \textbf{Integrated framework coordinating goal, semantic scope, resource envelope, physical--digital interaction, acceptance evidence, and lifecycle governance.} & \textbf{GOST links the six components through explicit dependencies. Changes propagate to affected model scope, placement, synchronization, acceptance, and lifecycle records; the resulting configuration is recomputed and validated before versioned reconfiguration, migration, or re-instantiation.} \\
\bottomrule
\end{tabularx}
\end{table*}
    
	\subsubsection{Goal-Oriented Principle} This layer represents the core execution phase that embodies GOST's goal-oriented and on-demand construction principles. Rather than pre-building a comprehensive twin, it dynamically instantiates a lightweight GOST instance tailored for a specific task when needed. The core idea is to start from business requirements and clearly define the twin's ``goal.''

    The goal-oriented principle delineates modeling scope and required precision. A network management objective may produce a DTN-like instance, whereas a structural inspection task may legitimately require a high-fidelity model. GOST does not rank either representation in isolation; it requires the chosen representation to be justified by the framework's joint design logic.
    
	
	Beyond delineating model scope, the goal-oriented principle emphasizes task comprehension, reasoning backward from goals to requirements \cite{10902595}. Instead of the intuitive approach of achieving goals directly through resource allocation, goal-oriented modeling converts ``soft goals'' into ``hard metrics'' \cite{chang2022kid}. This involves translating ambiguous task objectives (e.g., a UAV completing a search in a given area) into specific SAGSIN metric requirements (e.g., swarm endurance $\geq$ 20 minutes, satellite-UAV transmission rate $\geq$ 10 Mbps), thereby providing guidance for the optimization algorithms surrounding GOST.

In summary, GOST is defined as:

\begin{quote}
\textit{A SAGSIN twinning framework in which a mission goal, semantic scope, resource envelope, physical--digital interaction policy, acceptance evidence, and lifecycle governance form six interdependent components. It instantiates and updates task-relevant models over shared semantic foundations, and it retains bidirectional interaction only at the granularity and rate jointly determined by these components.}
\end{quote}

These six components are coordination variables rather than enabling technologies: goal defines the outcome and metric; semantic scope selects the relevant cross-domain entities and relations; resources bound communication, computation, and deployment; interaction policy selects exchange direction, granularity, and rate; evidence defines pre-action checks; and governance preserves version, provenance, and retirement records. The three GOST layers and Section IV technologies implement these decisions rather than replace them.


Table \ref{tab:dt_vs_st} makes the boundary explicit. Modern DT families supply essential components of GOST, and substantial overlap is expected. The distinction is compositional: adding an ontology, a semantic encoder, or a compressed model to a DT is insufficient unless task relevance governs scope and interaction, resource feasibility governs deployment, and evidence plus provenance governs acceptance and lifecycle transitions. Section V develops the corresponding evaluation criteria.

\subsection{Lessons Learned}
Table \ref{tab:new_twinning_paradigms_en_vcenter_4col} shows how GOST-related work combines a machine-readable knowledge blueprint, data-driven spatiotemporal and causal inference, and goal-derived lightweight instances. Together, these layers bridge heterogeneous devices, expose useful relationships in sparse data, and translate ambiguous objectives into actionable metrics while reducing communication and computation demands \cite{rico2023context,karabulut2024ontologies,Men2025Inter,zhu2021knowledge,thomas2023causal,kim2022data,9491087,raja2024towards}.

\section{Construction and Deployment of GOST} \label{section iv}

\begin{figure*}[t]
    \centering
    \includegraphics[width=0.98\textwidth]{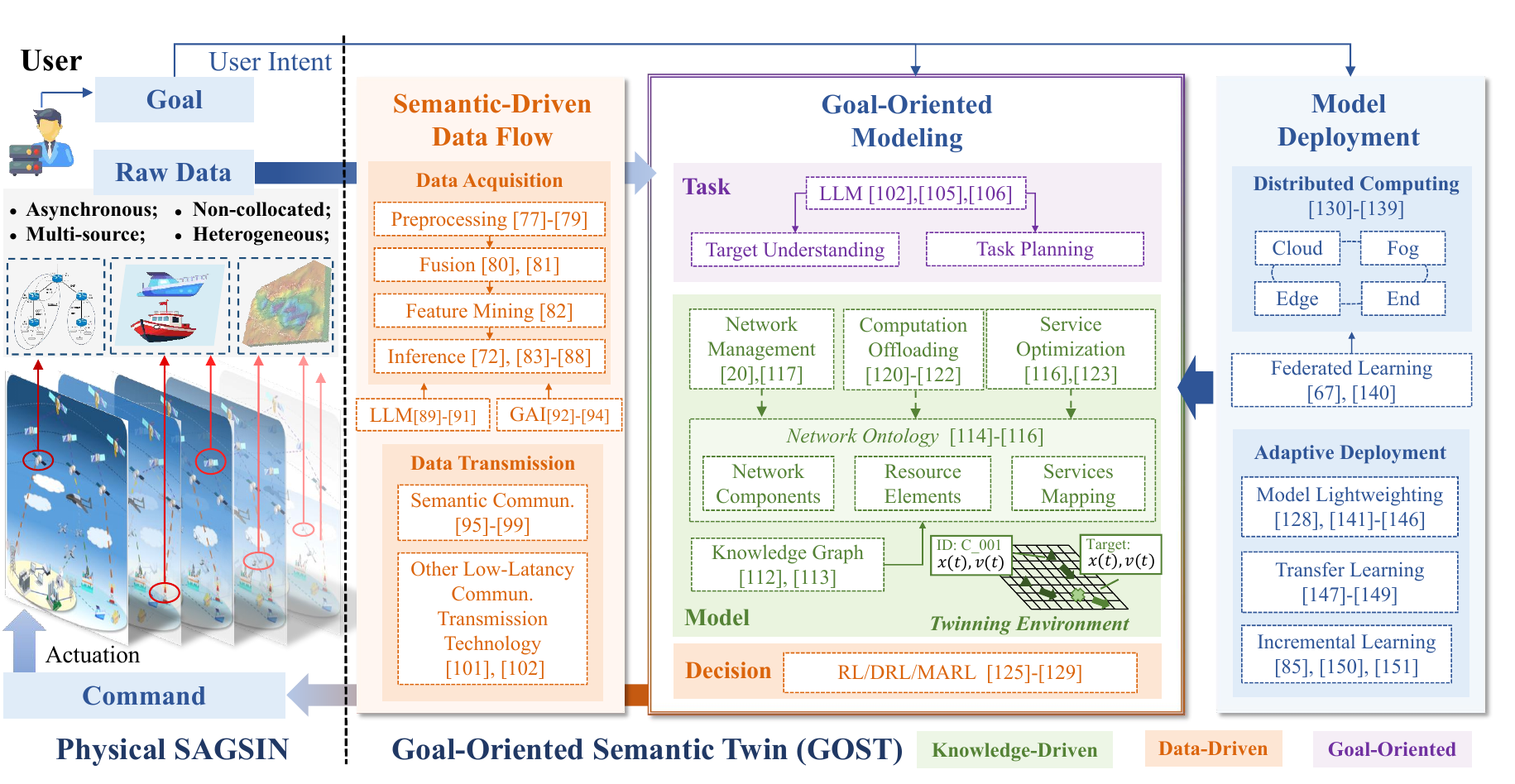} 
    \caption{The scope of GOST encompasses two main phases: the construction phase, which includes information acquisition, transmission, modeling, and decision-making; and the deployment phase, which covers adaptive deployment and distributed computing. Through semantic-driven data acquisition and transmission, goal-oriented network modeling, and distributed adaptive deployment, GOST enables precise perception, efficient computation, and dynamic adaptation within SAGSIN. References for key technologies in this section are listed in the figure.
}
    \label{tech_construction}
\end{figure*}

In the DT field, a widely accepted twinning framework is the 5D architecture proposed by Tao \textit{et al.} \cite{DT_Tao1}, encompassing physical entities, virtual entities, data, communication, and services. Similar to \cite{DT_Tao1}, this section is guided by Fig. \ref{tech_construction} and introduces semantic-driven data acquisition and transmission, goal-oriented modeling, and distributed adaptive deployment technologies to support the GOST framework within SAGSIN.

\subsection{Semantic-Driven Data Acquisition and Transmission}
Constructing GOST begins with establishing a timely and semantically enriched data flow between physical entities (PEs) and their digital counterparts. Achieving this necessitates not only semantic-driven data acquisition mechanisms, but also communication networks that support real-time synchronization\cite{Ouedraogo}. 

\subsubsection{Data Acquisition} To achieve such semantic-driven data acquisition, the massive, heterogeneous, and dynamic sensor data must first undergo preprocessing and quality assessment, enhancing the quality of datasets to ensure the accuracy of semantic extraction in GOST. In \cite{Taleb}, data cleansing is performed by evaluating data quality, where low-value information is discarded to avoid bandwidth wastage. Building on this foundation, the multimodal and high-dynamic data in SAGSIN also relies on a stable and scalable data management architecture \cite{Singh2020DataMF}, with related technologies including deep learning models such as convolutional neural networks \cite{LI2022167}.

After acquiring high-quality datasets, GOST entails the fusion and feature extraction of multi-source sensory data in SAGSIN, enabling data compression and precise perception of PEs. Common fusion techniques include traditional methods like signal processing and statistical approaches, as well as ML-based methods such as classification, regression, unsupervised learning, dimensionality reduction, and statistical inference\cite{Macias}. By effectively integrating multi-source perceptual data, GOST can identify key information within physical environments, thereby adapting in real-time to environmental and operational changes \cite{Zheng_Liu}. Feature extraction further reduces data dimensionality, as demonstrated by Wang \textit{et al.} \cite{Wang2023Intelligent}, who select a feature subset to reduce the dimensionality of high-traffic IoT data.

A key application of feature extraction in GOST is data inference and generation, in which unknown states or missing information are inferred from patterns in existing data. Depending on the characteristics of the data, inference typically unfolds along three complementary dimensions:

\begin{itemize}

\item \textbf{Temporal inference} predicts future states by time-series methods such as the maximum mean discrepancy algorithm \cite{11036541} or neural networks including RNN, LSTM \cite{11154046}, and GRU \cite{Jiahang_Xie}.

\item \textbf{Spatial inference} extracts and infers features from spatially distributed objects (e.g., networks or 3D scenes) through GNNs and CNNs to capture latent node relationships \cite{Xixi_Zhu}.

\item \textbf{Causal inference} identifies abstract relationships among variables by constructing causal graphs using regression, clustering \cite{Min}, or DNN-based approaches \cite{thomas2023causal}, enabling ``inferring effects from causes.''

\end{itemize}

By integrating these complementary perspectives, complex neural networks---such as convLSTM---can be constructed to perceive and extract multidimensional features, thereby enabling richer semantic perception in GOST \cite{deng_digital_2021}.

Beyond traditional data analysis methods, LLMs and other GAI technologies offer new capabilities for multimodal data processing in GOST systems \cite{Hong}. The authors in \cite{chai2024generative} and \cite{sivaroopan2024netdiffus} leverage GAI and diffusion models, respectively, to generate high-fidelity network samples for use by subsequent learning algorithms. Zhang \textit{et al.} \cite{Zhang2024GenerativeAA} focus on satellite communication networks, implementing an interactive modeling process using LLMs and employing retrieval-augmented generation (RAG) to extract satellite expertise that supports mathematical modeling. Moreover, to overcome data scarcity caused by environmental instability, GAI like VAE \cite{Litong_Zhang} and GAN \cite{Weiliang_He} can generate large amounts of high-quality data by synthesizing samples that preserve real-world statistical distributions \cite{Mannone, li2024automatic}.

\subsubsection{Data Transmission} The communication encompassed by GOST must achieve compressed data transmission with high timeliness over SAGSIN.

In contrast to traditional methods that rely on scheduling, modulation, and coding to achieve low-latency communication \cite{Bin_Li, Deng2024OptimizingAO}, GOST, with its characteristics at the knowledge-based and data-driven layers, is inherently well-suited for semantic communication within SAGSIN. This approach extracts semantics from signals at the transmitter for encoding and reconstructs the original information through semantic recovery at the receiver. This process reduces transmission bandwidth requirements \cite{Okegbile2025FLeSAF} and enhances information timeliness.

Semantic communication fundamentally leverages advanced AI and DL technologies to realize the paradigm of ``\textit{understand first, then transmit}.'' For example, \cite{Avi} designs an intelligent semantic communication system where the transmitter employs a domain-specific mobile segment anything model (MSAM) for semantic perception and extraction, while the receiver utilizes GAN to achieve high-quality data reconstruction under varying SNR. Besides \cite{Avi}, technologies such as DRL, GAI, and JSCC have also been widely applied to semantic adaptive sampling and extraction, and to mitigating channel fading \cite{Wanting_Yang}\cite{Weihan_Zhang}. 

Beyond understanding the transmitted information, the goal-oriented approach of transmitting only task-relevant information has also demonstrated promising results. In \cite{Talli}, the image compression and reconstruction are achieved via VAE, while employing DRL reward functions for joint optimization to focus compression on task-critical features. Chen \textit{et al.} \cite{11016707} applied semantic communication technology for sampling and reconstruction in a robotic arm, achieving precise on-demand twinning models through feature selection and DRL algorithms.

Semantic compression also concentrates risk: a small perturbation to a task-critical latent feature can change an action even when bit error rate remains low. An end-to-end semantic threat analysis for GOST should cover four connected attack surfaces: (1) sensor and training-data poisoning; (2) perturbations of task-critical features at the semantic encoder or wireless channel; (3) ontology/KG manipulation; and (4) compromised inference rules \cite{li2023data,gipivskis2023impact}. Protection spans the same chain. The transmitter estimates semantic uncertainty and sends redundant or raw evidence for high-impact concepts; the receiver checks signatures, ontology constraints, causal and physical consistency, and cross-modal agreement before accepting an update. Adversarial training and robust encoding reduce sensitivity, while a confidence or constraint violation triggers a fail-safe transition to verified raw telemetry or local control. Evaluation should report attack success rate, task-utility loss, semantic distortion, false acceptance/rejection, and fallback latency rather than communication errors alone.

\subsection{Goal-Oriented Modeling}

\subsubsection{The Task Analysis Process} As the core principle of GOST, the primary method in GOST modeling involves understanding objectives and transforming them into modelable, optimizable, and closed-loop metrics. As discussed above, goal-oriented modeling requires defining the scope, granularity, and evaluation of the twinning system based on goal requirements \cite{11016707}. However, the vast state space in SAGSIN makes it challenging for manual or simple DL models to capture key network elements. LLMs, with their powerful knowledge reasoning capabilities, are emerging as a potential key technology for task understanding and planning, enabling more intelligent and adaptive modeling of GOST in SAGSIN's complex tasks. Liu \textit{et al.} \cite{liu_delta_2024} propose an LLM-based robot task planning method that converts environmental topology into actionable knowledge, decomposing long-term goals into a sequence of autoregressive sub-goals. Zhang \textit{et al.} \cite{zhang_goal-guided_2025} design a target generator and policy planner based on LLM within the RL training environment to guide orderly exploration in RL, which also provides a viable solution for model training in complex SAGSIN environments.

Research on multi-objective optimization in SAGSIN shows a gradual shift away from representing heterogeneous requirements through a single weighted reward. Scalarization remains common, but its operating point depends strongly on preset weights and may obscure trade-offs among delay, coverage, energy, reliability, and throughput \cite{cui2021multiobjective,tu2024priority}. Recent studies increasingly keep service requirements visible while optimization proceeds: constrained multi-agent learning has been used to reduce routing delay while satisfying energy and packet-loss requirements \cite{lyu2024dynamic}, and priority-aware SAGIN slicing distinguishes service classes while jointly optimizing throughput, delay, and coverage \cite{tu2024priority}. Pareto analysis complements these formulations by presenting a set of efficient operating points instead of concealing the trade-off inside one scalar score \cite{cui2021multiobjective}. For GOST-oriented SAGSIN analysis, this literature supports retaining safety and physical feasibility as explicit constraints, using priority information to distinguish mission-critical from routine services, and comparing performance objectives only within the feasible region. Multi-stakeholder operation can therefore be organized as a two-level process: each agent exposes its minimum QoS, resource demand, and declared mission priority, and a GOST coordinator uses constraint-aware, priority-aware, and Pareto methods to generate feasible allocations of shared link and computing resources \cite{cui2021multiobjective,tu2024priority,lyu2024dynamic}. If one UAV requests a low-latency control update while another requests a high-accuracy sensing update, either the mission owner or a coordinating authority jointly appointed through a pre-agreed cross-operator policy selects or overrides the final allocation according to the declared task priorities, while GOST enforces safety constraints and records the decision.

\subsubsection{The SAGSIN Modeling Process} Once precise modeling objectives are set, GOST requires adequate models to describe the state and characteristics of the specific set of nodes under consideration in SAGSIN. In SAGSIN, ontologies enable cross-domain heterogeneous data semantic extraction and fusion through unified knowledge representation, ensuring semantic alignment and model accuracy, while also identifying key information and interrelationships among different data systems to streamline model migration, collaboration, and validation processes \cite{Han_Liu,Hoebert,lu2021cognitive}.

When constructing an ontology, a knowledge graph is a key technology for knowledge representation, modeling entities as nodes and relationships as edges \cite{Fangzhou_He}, thereby expressing complex, multi-level system relationships through flexible graph structures \cite{Tamasauskaite}.

\begin{figure}[t!]
    \centering
    \includegraphics[width=0.45\textwidth]{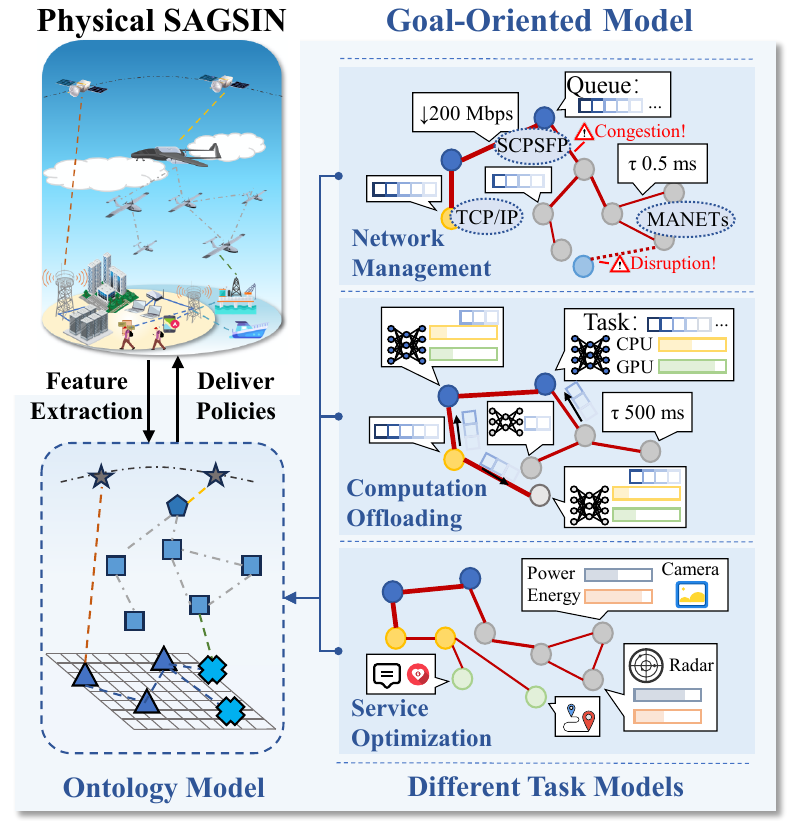} 
    \caption{Ontology for SAGSIN aimed at network management, computation offloading, and service optimization. For different goals, GOST allows distinct scopes and attributes for modeling, enabling evaluation and optimization across multiple dimensions \cite{tao2024wireless}.}
    \label{tech_net}
\end{figure}

Corresponding to the ontology method's modeling of concepts, attributes, and relationships, SAGSIN's network ontology requires three core elements: \textbf{network components}, \textbf{resource elements}, and \textbf{service mappings}. Network components include digital mappings of physical network nodes such as terminal devices, routers, base stations, and edge servers, with standardized node interaction functions, protocols, and interfaces \cite{campolo_network_2024}. Resource elements manage communication, computing, and service resources across nodes and links under specific objectives, supporting network resource scheduling \cite{Yejun}. Service mappings serve as the key to transforming ontology into goals by establishing connections between components, resources, and service capabilities \cite{yu_attention-based_2024}. Depending on different objectives, these network ontology elements may manifest in various forms, which we categorize into three types: network management, computation offloading, and service optimization, as illustrated in Fig. \ref{tech_net}.

\begin{itemize}

    \item Ontologies for network management require modeling communication environments, link quality, and node states. Key applications include ray-tracing for channel modeling \cite{alkhateeb2023real}, traffic simulations \cite{Yinzhi_Lu}, and virtual networks for lifecycle management \cite{10669844} and resource orchestration \cite{10107755}. Leveraging these states, Huynh \textit{et al.} \cite{van2022edge} optimize URLLC resource allocation, while Zhang \textit{et al.} \cite{zhang_mobility-aware_2024} utilize behavior prediction to improve edge service scheduling and latency.


    \item Ontologies for computation offloading focus on node metrics, QoS, and task queues, alongside predicting device behavior \cite{li_aoi-aware_2024}. As 6G and MEC mature, cloud-twin and edge-twin technologies now support flexible offloading \cite{khan_digital-twin-enabled_2022}. Liu \textit{et al.} \cite{liu_digital-twin-assisted_2022} develop an MEC twinning system integrating blockchain for intelligent server selection.
    

    \item Service-oriented ontologies emphasize GOST objective management by tracking specific resource distributions. In social networks, ontologies model user environments and preferences to foster connections \cite{chukhno_placement_2022}. For the Metaverse, they optimize communication capabilities alongside content parameters like resolution \cite{yu_attention-based_2024}.
\end{itemize}

\subsubsection{The Decision-Making Process} This process transforms precise states into effective control actions for physical network optimization. DL methods leverage powerful neural networks to achieve flexible and efficient state-to-action mapping. For simple data feature-based tasks like image recognition \cite{Baoxia_Du} and time series prediction \cite{deng_digital_2021}, approaches such as CNN and RNN introduced earlier can be employed by designing appropriate optimization objectives and neural network architectures. 

In decision-making and control tasks, DRL demonstrates exceptional adaptability in SAGSIN's complex environments \cite{Yihong_Tao}. Through continuous interaction with the environment, DRL employs reward mechanisms to guide agents in learning optimal strategies for complex scenarios, significantly improving efficiency in tasks such as network resource allocation and UAV path planning \cite{Xin_Tang}. 

Building upon DRL agents, multi-agent systems leverage distributed collaboration and game theory mechanisms to achieve global knowledge sharing and federated evolution, significantly enhancing system scalability and fault tolerance \cite{Zhou2024HierarchicalDC}. In edge DT optimization for satellite-ground networks, multi-agent systems can be used to simulate autonomous decision-making in scenarios involving multiple terminals, sensors, and other devices, thereby enabling reliable sampling and efficient transmission \cite{Yihong_Tao}\cite{Ruah}. With the scaling and intelligent development of SAGSIN, multi-agent networks will substantially enhance GOST's adaptability and decision-making capabilities in complex environments and tasks.

\subsection{Distributed and Adaptive Deployment}

Following the goal-oriented modeling mechanisms discussed above, this subsection shifts the focus to how GOST is effectively deployed in SAGSIN environments. Unlike conventional digital twin systems, GOST must operate across heterogeneous nodes with uneven computing capabilities, rapidly changing network topologies, and mission-driven performance requirements. To overcome these challenges, we review three enabling pillars: distributed architectures, model lightweighting, and adaptive learning mechanisms.

\subsubsection{Distributed Computing} As SAGSIN scales, a single GOST model can no longer meet the management demands of such complex systems, where centralized computing and control architectures struggle to satisfy multi-tasking and real-time requirements \cite{wang2025multi}. In light of these limitations, distributed GOST leverages a cloud-edge-end collaborative architecture and parallel computing technology to achieve dynamic task allocation and optimization \cite{Jiang,Xu,Hofmann}, enabling collaborative development, intelligent integration, and monitoring of multiple GOST models \cite{Yang_Liu}. Specifically, the cloud infrastructure serves as a central hub for advanced intelligence, enabling multi-task collaborative reasoning through model aggregation\cite{Burger} and fostering knowledge sharing for semantic graph construction \cite{Wang}. To further refine resource scheduling within this framework, fog computing is introduced as an intermediate layer to alleviate computational loads\cite{Mi}. This extended cloud-fog-edge-end collaboration facilitates efficient resource utilization and PCCA optimization\cite{Yihong_Tao,11063292}, enabling multi-level distributed decision-making strategies to ensure robust control in dynamic environments\cite{Pan}.

\begin{figure}[t!]
    \centering
    \includegraphics[width=0.485\textwidth]{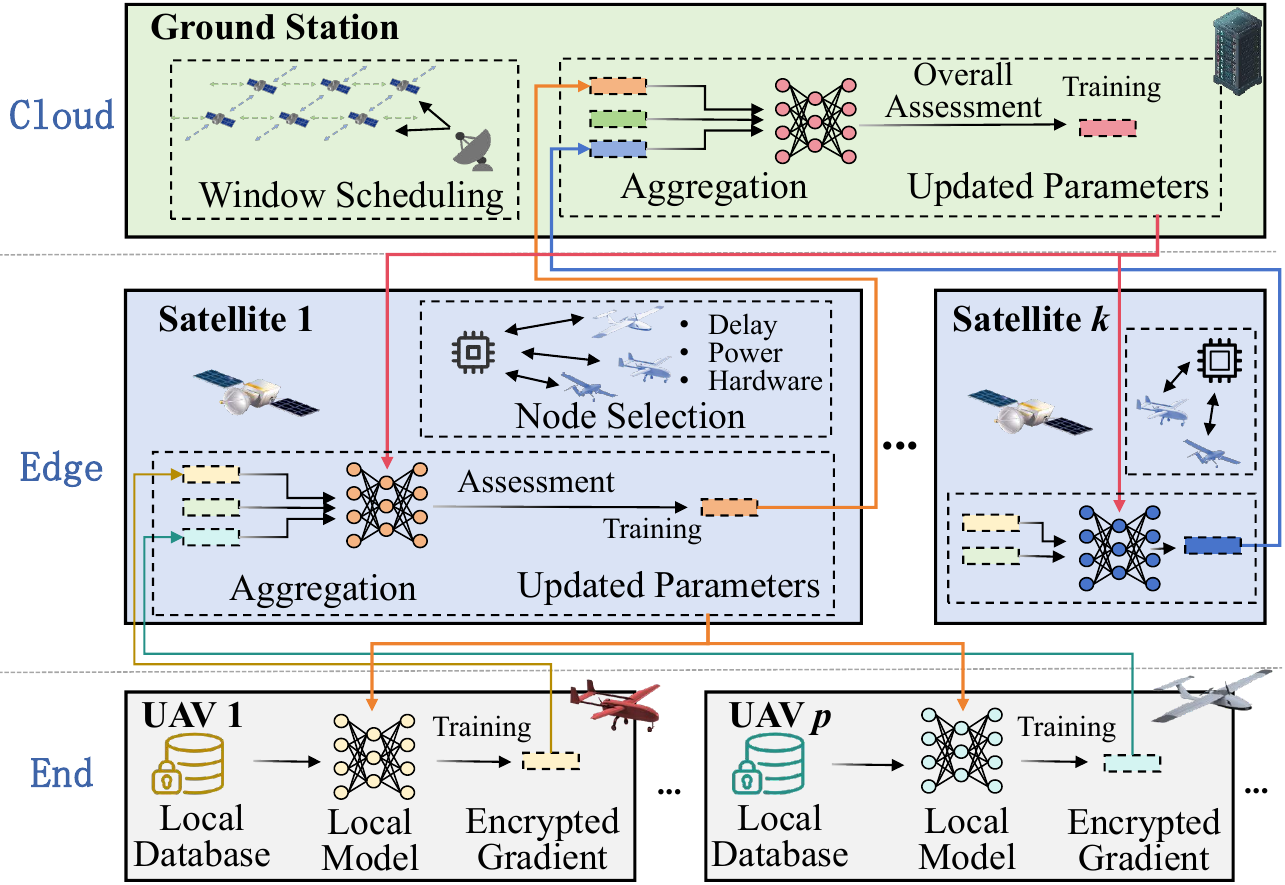} 
    \caption{An illustration of cloud-edge-end collaborative FL in SAGSIN, where the GCS, satellites, and UAVs serve as cloud, edge, and end nodes respectively. As a distributed ML paradigm, FL enables multiple devices to collaboratively train a global model without sharing their local private data \cite{lu2020communication}. 
}
    \label{fig_FL}
\end{figure}

Federated learning (FL) represents an essential enabler for distributed GOST deployment, as it allows multiple heterogeneous nodes to collaboratively train and update models without centralized data aggregation. A cloud-edge-end collaborative FL framework in a ground-satellite-UAV network is presented in Fig. \ref{fig_FL} to demonstrate how GOST utilizes FL within SAGSIN. FL has been widely applied for model aggregation, optimization, and updates \cite{lu2020communication}, providing a reliable learning framework for large-scale, highly heterogeneous SAGSIN environments. For example, in \cite{9491087} and \cite{Yunlong_Lu_L}, FL is employed to optimize access strategies for mobile edge users and migration strategies for DT.

\subsubsection{Adaptive Deployment} The severe resource constraints of edge nodes (e.g., satellites, UAVs) prevent the direct deployment of high-precision models. This necessitates model lightweighting techniques that reduce model size and inference latency while maintaining performance, enabling efficient operation on resource-limited devices \cite{Wei_Song,Gupta}. Technically, model lightweighting in GOST is achieved through complementary strategies---knowledge distillation \cite{Xiong} for transferring high-level representations, pruning \cite{Zhou2024HierarchicalDC} for structural sparsification, and quantization \cite{Rehman} for reducing numerical precision---each reducing computational cost from a different perspective. A notable implementation is demonstrated by Zhou \textit{et al.} \cite{Xiaokang_Zhou}, who study edge model deployment and global model aggregation in an end-edge-cloud multi-level DT. By combining model pruning with federated bidirectional distillation, their approach significantly reduces communication overhead and computing load. 

LLMs can provide strong reasoning capabilities for the PCCA loop, but their deployment has been traditionally constrained by massive parameter scales \cite{Hadish}, which necessitates lightweighting techniques such as split learning/inference, parameter-efficient fine-tuning, shared-parameter inference, and collaboration between small and large language models, thereby enabling end-edge large model inference \cite{lin2025pushing}.

Beyond static deployment feasibility, the highly dynamic nature of SAGSIN requires GOST models to support rapid parameter transfer and consistency maintenance during node handovers or topological changes\cite{10886920}. This requires learning strategies that support both swift cross-domain adjustment and ongoing in-domain evolution.

Transfer learning (TL) provides an efficient solution for cross-domain model adaptation by reusing knowledge from related tasks, thereby avoiding the computational burden of repeated training \cite{9522071}. For instance, \cite{Yan_Xu} utilizes deep TL to transfer pre-trained models from virtual to physical spaces, achieving timely and reliable predictive maintenance.

Incremental learning complements TL by enabling continual in-domain updates as the environment evolves, often through strategies such as meta-learning and lifelong learning \cite{Finn, qyf_lifelong}. Incremental learning can be further enhanced by leveraging cloud–edge collaborative architectures \cite{Jiahang_Xie} or federated learning frameworks \cite{Qian_Wang}, which enable GOST nodes in SAGSIN to offload continual learning tasks across distributed nodes for more efficient model evolution.

Lifecycle research treats accountability as a continuity problem: the virtual representation, its data, and its governing models must remain identifiable as they evolve. Systematic DT reviews identify lifecycle coverage and data ownership as persistent research gaps \cite{jones2020characterising}, while ontology-based lifecycle work models twin-data evolution and governance across multiple granularities \cite{9813407}. For distributed and short-lived GOST instances, these findings favor persistent identifiers and versioned provenance that link an action to the applicable goal, ontology/KG snapshot, input lineage, encoder and decision-model versions, deployment node, timestamps, uncertainty, output, and acceptance evidence. The record should follow creation, update, migration, rollback, and retirement so that a later version is not mistaken for the state that produced an earlier decision. Research on synchronized provable data possession shows that blockchain-backed mechanisms can verify time-state consistency and data integrity across distributed virtual spaces \cite{li2022synchronized}; such mechanisms are optional protection for shared records rather than a substitute for deciding what provenance must be retained. A complementary line uses formal specification and executable analysis to examine modeled DT behavior \cite{alhazmi2026formal}. Taken together, the literature points toward combining versioned lifecycle governance, tamper-evident provenance, and replay or formal analysis, while retention and disclosure remain subject to operator privacy and sovereignty policies.

\subsection{Lessons Learned}
GOST construction links data acquisition, goal-oriented modeling, and distributed adaptive deployment. Semantic centrality must be maintained across the data lifecycle so that heterogeneous nodes can exchange understandable information, while model placement and lightweighting adapt to communication, computing, and task conditions \cite{Avi,Wanting_Yang,Weihan_Zhang,Talli,Baoxia_Du,Wei_Song,Gupta,Xiong,Zhou2024HierarchicalDC,Rehman}.

\section{Beyond High Fidelity: A Multidimensional Evaluation Framework for Utility and Semantics}

\begin{figure*}[t]
	\centering
	\includegraphics[width=0.95\textwidth]{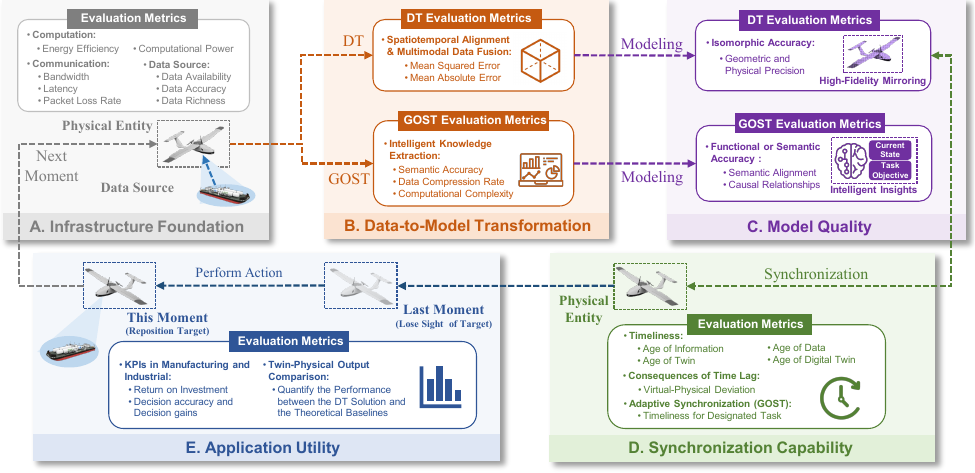}
	\caption{A five-dimensional evaluation framework for GOST aligned with the PCCA loop, encompassing infrastructure foundation, data-to-model transformation, model quality, synchronization capability, and application utility.}
	\label{evaluation_GOST}
\end{figure*}


As noted by Zhang \textit{et al.}, evaluating the correctness of a DT is pivotal to practical utility \cite{ZHANG2021151}. Fig. \ref{evaluation_GOST} illustrates five PCCA-aligned dimensions: infrastructure, data-to-model transformation, model quality, synchronization, and application utility. We add trustworthiness and governance as a cross-cutting plane rather than an isolated sixth stage, because security, conflict handling, and accountability must be tested at every transition. The plane uses semantic attack success and task utility loss for robustness; conflict resolution time, infeasible-goal detection, and constraint-violation rate for multi-goal coordination; and provenance coverage, audit completeness, integrity verification, and decision-replay success for governance. A GOST is acceptable only when both its task metric and the relevant cross-cutting thresholds are satisfied.

\subsection{Infrastructure Foundation: Computing Resources and Data Sources}

This dimension measures the resource ceiling imposed by hardware, communication, and sensing. Device-level indicators include FLOPS, memory and storage bandwidth, and energy consumption \cite{wu2025towards,tran2025network}; system-level indicators compare cloud and edge placement and may quantify architectural complexity, stability, and efficiency through network architecture entropy \cite{JuniperMEC2016,wang20236g}. Communication is evaluated through bandwidth/data rate, latency, packet loss, jitter, path loss, and SINR \cite{mihai2022digital,erceg1999empirically,kurt2017path,maccartney2013path}. Perception quality is evaluated through sensor-data availability, accuracy, and modality richness \cite{taleb2025big}.


\subsection{Data-to-Model Transformation: Processing and Abstraction}

This dimension evaluates the critical step of converting raw, heterogeneous data into structured and meaningful inputs for virtual model construction. In SAGSIN, the scale is typically large, involving numerous physical entities and a vast number of sensors, resulting in massive and complex data flows. Rasheed \textit{et al.} summarized the characteristics of modern data with the well-known ``4Vs'': volume, velocity, variety, and veracity \cite{rasheed2020digital}. Efficiently handling these data under constrained hardware and communication resources is key to realizing system performance.

Conventional DT targets exhaustive, lossless replicas. While fit-for-purpose and adaptive-fidelity strategies have relaxed this uniform ambition, GOST formalizes task-conditioned functional adequacy, tailoring twin content to the functional requirements of specific tasks. Thus, evaluation focuses on whether the system can efficiently and accurately integrate data from different sources, timestamps, and formats. 

The first capability is \emph{spatiotemporal alignment}, typically evaluated by quantifying the precision of aligning multi-source data in both temporal and spatial dimensions, which holds significant reference value for evaluating GOST in the multi-source SAGSIN environment. In quantitative terms, the literature frequently employs statistical metrics such as mean squared error (MSE), mean absolute error (MAE), and F1-score \cite{chang2024align,liu2025calf,li2025spatio}, alongside similarity measures like dynamic time warping (DTW)\cite{ZHAO2018171}. 

The second capability is \emph{multimodal data fusion}, which evaluates the ability to combine heterogeneous sensor data (e.g., vision, radar, temperature) to improve overall modeling accuracy and robustness. Contemporary works often employ paradigms such as unimodal vs. multimodal comparisons and robustness testing to assess performance under conditions of missing modalities, noise interference, or computational constraints \cite{liu2022multi,zhou2024multimodal,zhang2023digital}. Common metrics cover perception (mAP, IoU, HOTA), prediction and regression (MAE, RMSE, DTW), uncertainty (ECE, Brier score, NLL), as well as system-level latency and energy consumption.

Beyond the metrics already mentioned, this stage emphasizes uncovering underlying syntactic rules, identifying semantic associations among entities, and constructing lightweight yet knowledge-rich models \cite{11078840,11016707}. Accordingly, GOST additionally requires the following evaluation metrics:


\begin{itemize}
    \item \textbf{Evaluation of semantic tools:} Xue \textit{et al.} present a holistic KG evaluation framework focusing on accuracy, consistency, completeness, timeliness, and redundancy, employing both intrinsic indicators (such as the ratio of valid facts) and extrinsic indicators (such as downstream task gains) for quantitative analysis \cite{xue2022knowledge}.
    
	\item \textbf{Semantic accuracy/ambiguity:} Measures the precision of semantic information extraction \cite{xie2021deep,wang2021faier}. Metrics based on mutual information (MI) and entropy are commonly used to evaluate the accuracy of semantic reconstruction \cite{li2023image,Chen2022AMI,wang2025multi}. Existing GOST-related studies frequently employ MI, entropy, KL divergence, and cross-entropy loss as evaluation metrics \cite{11129865,ren2024digital}.
	\item \textbf{Data compression rate / semantic compression:} A key metric is the reduction in data size while preserving the semantic information required for tasks \cite{liu2023adaptable}. This enhances the usability of GOST within SAGSIN. By extracting semantic information from reconstructed messages, \cite{11016707} achieves over a 50\% reduction in communication load.
	\item \textbf{Computational complexity:} Algorithmic complexity impacts power consumption and timeliness in bidirectional interactions and is generally estimated through algorithmic time and space complexity analysis or empirically measured. \cite{Apostolakis2023DigitalTF} measures the average inference time of DT neural network models and compares it with the time required for purely physical simulation.
\end{itemize}

\subsection{Model Quality: Fidelity and Equivalence}

This dimension evaluates the twinning model itself by measuring how well it represents its physical counterpart. Here, ``accuracy'' is redefined. A DT is evaluated based on \emph{isomorphic accuracy} (the degree to which the twin resembles the real object in appearance and behavior), while GOST is evaluated based on \emph{functional or semantic accuracy}. A GOST may bear no physical resemblance to the real object; as long as it can accurately predict outcomes or meaningfully capture the system state, it can still be considered ``highly accurate.''

Grieves' original vision for DT required extremely high-fidelity representation, extending even to atomic-level fidelity \cite{sharma2022digital}. Indicators included dimensional tolerances, material property deviations, and the consistency of simulation results with known physical laws.

Subsequent research has recognized that Grieves' ultimate vision is currently unattainable, and the focus has shifted toward balancing fidelity against cost \cite{kober2022digital}. Several studies \cite{clement2017internet,kober2022digital} suggest that fidelity should be tailored to the application scenario, considering resource and contextual constraints. 
In this context, many evaluation studies design key indicators as weighted functions or nonlinear transformations that jointly consider fidelity and external parameters \cite{Zhou2024HierarchicalDC}.

In contrast, GOST aims to construct a model that accurately reflects the fundamental characteristics, operational rules, and semantic linkage information of physical systems. Although there is no unified consensus on evaluation dimensions and metrics, relevant fields provide useful insights.

Getu \textit{et al.} survey various semantic-oriented metrics applicable to GOST evaluation, including inference capability metrics (quantifying the divergence between teacher and apprentice models), real-time reconstruction error, and cost of actuation error \cite{10107602}. Similarly, Feinglass \textit{et al.} introduce semantic proposal alikeness rating using concept similarity (SPARCS), which evaluates semantic alignment between twin outputs and real system states at the conceptual level \cite{feinglass2021smurf}. Thomas \textit{et al.} quantify whether the established model correctly captures causal relationships by measuring the NMSE between virtual and real system transitions \cite{thomas2023causal}. To assess the extent to which a twin system has ``understood'' system states, Kim \textit{et al.} propose testing with concept activation vectors, which measures a neural network's sensitivity to specific concepts during class recognition \cite{kim2018interpretability}. Bau \textit{et al.} further introduce the network dissection method, quantifying the interpretability of CNN latent representations by scoring each convolutional unit against semantic concepts \cite{bau2017network}. 

At the goal-oriented level, GOST often compresses larger, more complex models (such as full physical simulation models) into lightweight, data-driven approximations. Evaluation of such lightweight models typically compares the compressed model with the source model across task performance (e.g., accuracy, F1, mAP, and RMSE), distribution similarity (e.g., KL divergence, cosine similarity), and resource efficiency (e.g., parameter count, FLOPs, and inference latency) \cite{Xiaokang_Zhou,10549920,Xiong}.

\subsection{Synchronization Capability: Latency and Consistency}

Synchronization is the process of maintaining temporal alignment between the physical and virtual states \cite{cakir2023synchronize}, that is, the ability of the twin system to stay synchronized with its physical counterpart. This dimension specifically evaluates such dynamic consistency, measuring the extent to which the virtual world can track physical changes in real time.

In assessing the timeliness of DT and GOST, the most direct method is to quantify the ``freshness'' of data or states within the twin. Much of this work leverages concepts from communication theory---particularly AoI, defined as the time elapsed since the generation of the most recently received information packet. Building on AoI, several works \cite{duran2023age,shu2022age,noroozi2025age} have proposed concepts such as age of twin (AoT) and age of data (AoD). These measures typically center on the time elapsed since the last twin state update; the smaller the value, the fresher the twin model, and the lower the uncertainty in its representation of the physical entity. Their advantage lies in unifying diverse delays and packet losses during DT/GOST updates into a single simple parameter, offering strong generality and broad applicability.

Beyond directly measuring elapsed time, some works measure the consequences of time lag---namely, the deviation between virtual and physical states. This approach is more goal-oriented, as it directly quantifies the impact of synchronization delay on accuracy, essentially a continuous temporal measurement of fidelity and equivalence \cite{lugaresi2023online}. Cardoso \textit{et al.} introduce a dedicated synchronization measure that outputs a percentage value representing the relative distance and synchronization precision between the virtual entity, the physical entity, and dynamic targets (e.g., planned trajectories), providing a quantitative evaluation standard for kinematic synchronization \cite{Cardoso2025ModelingEA}. When reviewing DTN, Duran \textit{et al.} summarize indicators such as data handling and delivery rate for specific services \cite{11261676}. Such approaches are frequently employed in lifecycle management of physical entities. While more accurate for specific use cases, they are less generalizable.

\subsection{Application Utility: Verification and Task Outcomes}

A DT or GOST is a tool for achieving specific objectives, not an end in itself; its value is therefore best evaluated through its final task outcomes. 

By applying KPIs, DT or GOST can be evaluated in a standardized manner. In the manufacturing and industrial sectors, Tang \textit{et al.} propose a comprehensive evaluation framework for a high-speed railway DT platform, covering performance, user experience, and economic benefits, serving as a reference for domain-specific evaluations \cite{tang2025evaluation}. For SAGSIN, KPIs can be designed around specific tasks, including reduced decision-making time, beam prediction accuracy \cite{jiang2023digital}, and enhanced correlation between GOST and the physical environment.

In academia, the evaluation is often conducted by comparing the outputs or decisions of DT or GOST with returns from the real world or with baseline approaches \cite{lugaresi2023online}, similar to reinforcement learning paradigms where agents are iteratively assessed. Cao \textit{et al.} compare results from DT-driven optimization algorithms with known optimal solutions and use regret values to quantify the performance loss between the DT-derived solution and the optimal one \cite{cao2021simulation}. Zhou \textit{et al.} evaluate DT by inputting simulation data following specific statistical distributions into the virtual environment and comparing the statistical properties of the output against theoretical baselines \cite{Zhou2024HierarchicalDC}. In SAGSIN, such evaluation methods can also extend to comparisons between different GOST models, or benchmarking against cloud-based DTs using historical data. 

\begin{table*}[htbp]
    \centering
    \caption{Selected Key References for the Evaluation of GOST in SAGSIN}
    \label{tab:gost_sagin_eval_refs_en}
    \small 
    \begin{tabularx}{\textwidth}{
			>{\RaggedRight}p{3cm} 
			>{\RaggedRight}p{1.1cm}
			>{\RaggedRight}X
            >{\RaggedRight}p{10.5cm}
		}
        \toprule
        \textbf{Evaluation dimension} & \textbf{Ref.} & \textbf{Specific Aspect} & \textbf{Core Contribution / Relevance} \\
        \midrule
        
        Infrastructure Foundation & \cite{JuniperMEC2016} & Hardware Architecture & Compares cloud vs. edge computing on latency and energy efficiency, providing a basis for twin system deployment decisions. \\
        \cmidrule(r){2-4}
         & \cite{wang20236g} & Network Architecture & Proposes the innovative ``network architecture entropy'' concept to holistically quantify the design quality of networks serving DT. \\
        \midrule
        
        Data-to-Model Transformation& \cite{rasheed2020digital} & Data Characteristics & Summarizes the ``4Vs'' (volume, velocity, variety, veracity) of data in modern twin systems. \\
        \cmidrule(r){2-4}
         & \cite{11016707} & Semantic Compression & Reduces communication load by over 50\% via extracting semantic information, demonstrating the lightweight potential of GOST. \\
        \midrule

        Model Quality & \cite{wang2024optimal} & Balancing Fidelity & Suggests that fidelity should be tailored to application, resources, and cost, representing a paradigm shift from ``maximum precision'' to ``fit for purpose.'' \\
        \cmidrule(r){2-4}
         & \cite{Xiaokang_Zhou} & Goal-Oriented Evaluation & Presents knowledge distillation as a representative method for model lightweighting, with evaluation criteria focusing on the balance between task performance and resource efficiency. \\
        \midrule

        Synchronization Capability & \cite{duran2023age} & State Freshness & Introduces a derivative of AoT to quantify the freshness of the twin state, integrating factors like delay and packet loss. \\
        \cmidrule(r){2-4}
         & \cite{ganguli2020digital} & Goal-Oriented Sync & Proposes the concept of multi-scale synchronization, emphasizing differentiated evaluation based on parameter sensitivity. \\
        \midrule

        Application Utility & \cite{cao2021simulation} & Optimization Evaluation & Uses ``Regret'' to quantify the performance gap between twin-driven optimization algorithms and the optimal solution. \\
        \cmidrule(r){2-4}
         & \cite{tang2025evaluation} & Industrial Framework & Proposes an evaluation framework for a high-speed railway DT, covering performance, user experience, and economic benefits, and serving as a model for domain-specific applications. \\
		\midrule
		Cross-Cutting Trust & \cite{gipivskis2023impact} & Semantic Robustness & Measures adversarial success and downstream task loss, exposing failures that bit-level metrics can miss. \\
		\cmidrule(r){2-4}
		 & \cite{jones2020characterising,9813407,li2022synchronized,alhazmi2026formal} & Provenance and Audit & Assesses persistent identity, versioned data/model lineage, tamper-evident records, replay support, and formally analyzable behavior across instance creation, update, migration, and retirement. \\
		\cmidrule(r){2-4}
		 & \cite{cui2021multiobjective,tu2024priority} & Multi-Goal Coordination & Reviews constraint-aware, priority-aware, and multi-objective methods for exposing conflicts and performance trade-offs in integrated networks. \\

        \bottomrule
    \end{tabularx}
\end{table*}

In terms of semantic metrics, Meng \textit{et al.} \cite{meng2024effectiveness} propose the synchronization cost of information (SCoI), which consists of two components: a task-specific state estimation cost and an actuation cost. In the context of GOST-empowered SAGSIN, the state estimation cost can be derived from understanding at the goal-oriented layer, while the actuation cost can stem from pre-defined models. Similar metrics include AoII, VoI, and UoI, all of which relate to the final utility of information \cite{maatouk2020age,lu2023semantics,zheng2020urgency}. 

\subsection{Lessons Learned}

Table \ref{tab:gost_sagin_eval_refs_en} summarizes a multidimensional evaluation framework spanning infrastructure, data-to-model transformation, model quality, synchronization, and application utility. The dimensions are interdependent: effective assessment must remain goal-oriented because optimizing one layer in isolation can undermine overall system performance \cite{11261676}.


\section{Case Study: a GOST System for Remote Multi-UAV Tracking}
\begin{figure*}[t]
	\centering
	\includegraphics[width=0.99\textwidth]{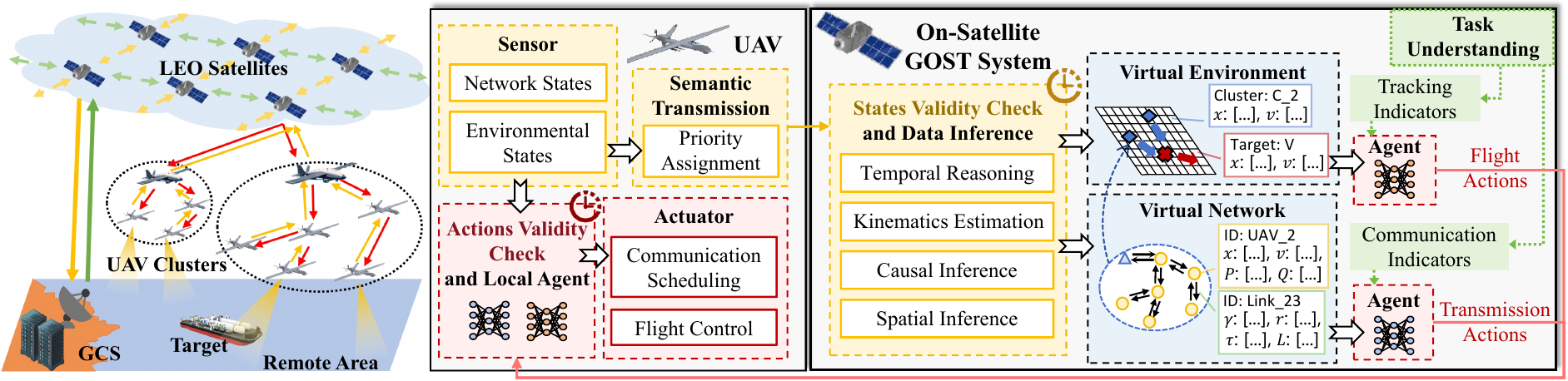}
	\caption{A GOST system for remote multi-UAV tracking. Based on an ontological model of satellite-UAV networks, it ensures the timeliness and reliability of GOST through semantic synchronization and transmission mechanisms, and guarantees the precise task execution through task parsing and MARL.}
	\label{case_sys}
\end{figure*}
\subsection{Task Scenarios and the GOST Architecture}
This section demonstrates the construction and application of a GOST system through a typical SAGSIN scenario: a remote multi-UAV collaborative target tracking task \cite{LTL_GOST_TMC}, as illustrated in Fig. \ref{case_sys}. The task area is set in a remote maritime region spanning 100 km $\times$ 100 km with a weak communication infrastructure. In this scenario, two UAV clusters (each consisting of 4--6 glider UAVs) patrol at 300-meter altitude, tracking moving sea targets with unknown maneuvering patterns. An LEO communication-computing integrated satellite constellation (operating at 550 km altitude) provides the sole backhaul link and edge computing services, while a ground control station (GCS) located 2000 km away serves as the cloud intelligence hub. Channel modeling for UAV-satellite links and inter-UAV communications adopts 3GPP TR38.811 NTN and IEEE 802.11n standards, respectively. This scenario highlights SAGSIN's characteristics of high dynamism, extreme heterogeneity, resource scarcity, and large-scale topological changes, requiring stringent real-time situational awareness and collaborative decision-making capabilities.

In this scenario, the GOST implementation for the satellite-UAV network strictly adheres to the three-layer architecture proposed in Section \ref{section iii}. The knowledge-based semantic layer instantiates the UAV network through a structured ontology. It maps physical entities---including network topology (represented as directed graphs), resource attributes (UAV mobility and channel quality), and service logic (UAV-satellite-ground interactions)---into a semantic knowledge base that defines information flow patterns. The data-driven semantic layer employs a spatiotemporal-causal inference engine integrating ARIMA for time compensation, UIF for motion estimation, GAT for spatial correlation, and DECI for causal discovery to maintain state continuity of the twinning system in the communication-constrained network, thereby enhancing its environmental awareness. The goal-oriented principle layer translates high-level task objectives into quantifiable ``effective collaborative tracking'' targets and decomposes them into optimizable metrics through LLM-based parsing, such as AoI $< 30$~ms, positioning error $< 10$~m, and specific safety constraints regarding tracking and collision risks.

The aforementioned GOST model is deployed on LEO satellites equipped with sufficient computing power. During the task, UAVs report their status and environmental conditions through the uplink according to a predefined strategy. The satellite then updates the GOST model, performs semantic inference and policy optimization, and issues collaborative commands via the downlink. The simulation platform is developed in Python and SimPy with a system synchronization cycle of 100~ms.

\subsection{GOST-Empowered Timely and Reliable Data Transmission}
\begin{figure}[htbp]
	\centering                 
	
	\begin{subfigure}{0.24\textwidth}
		\centering
		\includegraphics[width=\linewidth]{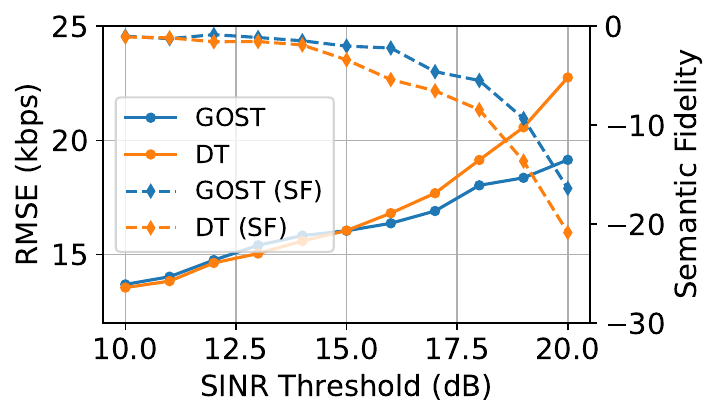}
		\caption{Data generation rate}
		\label{infer_delay}
	\end{subfigure}
	\hfill                     
	\begin{subfigure}{0.24\textwidth}
		\centering
		\includegraphics[width=\linewidth]{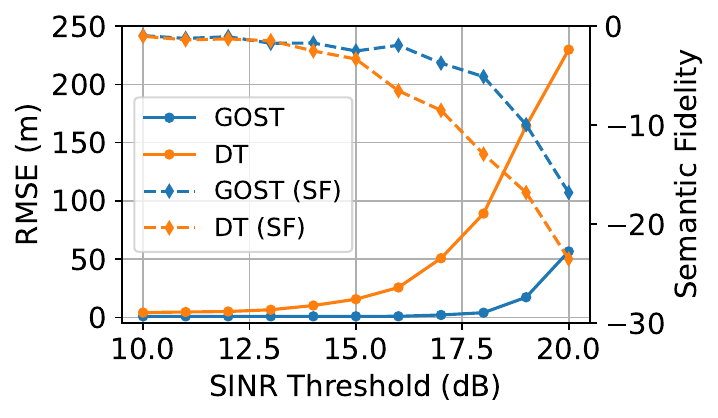}
		\caption{Estimated target position}
		\label{infer_move}
	\end{subfigure}
	
	\vspace{0.1cm}             
	
	\begin{subfigure}{0.24\textwidth}
		\centering
		\includegraphics[width=\linewidth]{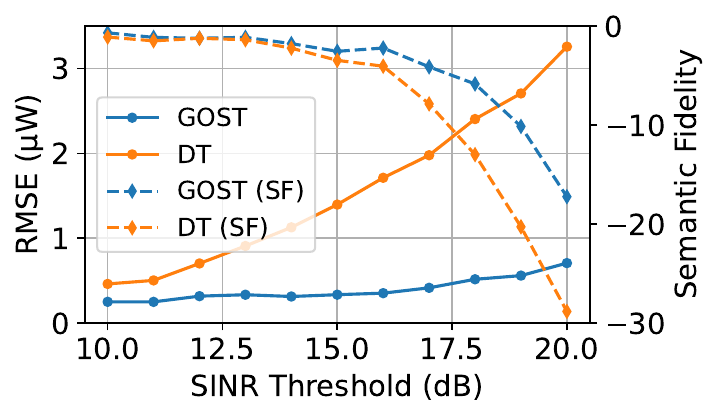}
		\caption{Interference power}
		\label{infer_signal}
	\end{subfigure}
	\hfill
	\begin{subfigure}{0.24\textwidth}
		\centering
		\includegraphics[width=\linewidth]{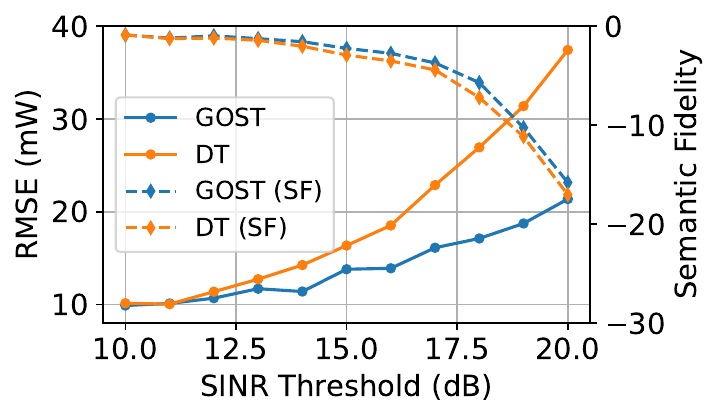}
		\caption{Power consumption}
		\label{infer_state}
	\end{subfigure}
	
	\captionsetup{subrefformat=parens}  
	\caption{Accuracy and semantic fidelity of typical status data in GOST and compressed DT, including data generation rate and estimated target position (temporal feature), interference power (spatial feature), and power consumption (causal feature).}
	\label{case_error}
\end{figure}
\begin{figure}[t!]
	\centering
	\includegraphics[width=0.45\textwidth]{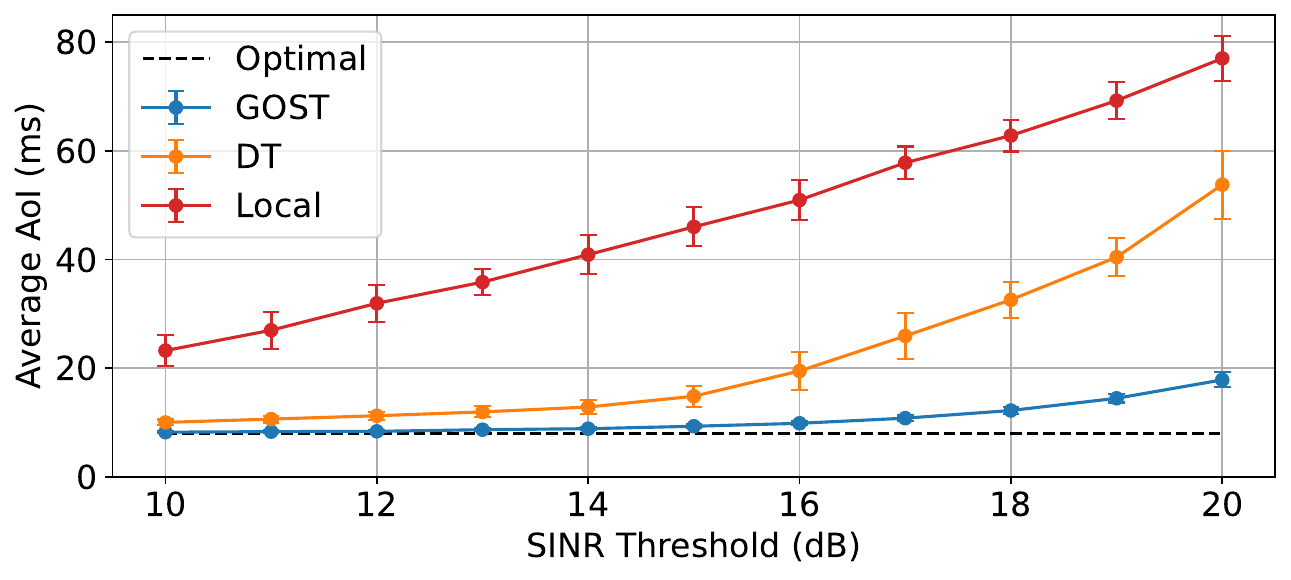}
	\caption{Communication scheduling performance of GOST, compressed DT, and local models, with the error bars representing the standard deviation.}
	\label{case_aoi}
\end{figure}

The GOST system employs an adaptive synchronization mechanism to collect network and environmental states through semantic-aware scheduling. It dynamically adjusts reporting priorities based on causal relationships between state data and inference accuracy, while optimizing transmission power and frequency bands using a communication scheduling agent trained by MADDPG. For comparison, we use a compressed DT baseline. Under the same bandwidth budget, this baseline applies channel-aware compression uniformly across all state variables to increase their update opportunities toward the delivery probability achieved by GOST. It retains global synchronization but lacks task-aware prioritization and semantic inference; lower bit depth therefore introduces accumulated quantization noise as channel quality deteriorates.

Fig. \ref{case_error} jointly presents inference accuracy (RMSE, left axis) and semantic fidelity (right axis) for temporal, kinematic, spatial, and causal state variables under different channel conditions. Semantic fidelity summarizes the coupled effect of single-sample estimation bias and channel-dependent distribution shift on downstream decision utility. As channel quality deteriorates, the compressed DT's quantization error increases substantially: its RMSE values for position and interference estimation become nearly four times those of GOST because uniform compression cannot preserve task-critical information. GOST instead compensates for packet loss through temporal, kinematic, spatial, and causal inference and maintains lower RMSE across all four variable categories. The right-axis results show higher semantic fidelity for GOST under nearly all evaluated channel conditions, indicating that its semantic compression preserves task-relevant structure rather than only reducing transmitted bits.

Fig. \ref{case_aoi} further evaluates communication scheduling through average AoI and includes compressed DT and a local model based on distributed games. GOST approaches the theoretical optimum by prioritizing key variables and inferring missing states, reducing AoI by 66\% relative to compressed DT. Compressed DT performs similarly when communication resources are ample, but its AoI deteriorates rapidly with substantial jitter when the SINR threshold exceeds approximately 15~dB because state-estimation errors prevent its scheduling model from reliably computing transmission actions.

\subsection{Goal-Oriented Decision-Making}
\begin{figure}[htbp]
	\centering                 
	
	\begin{subfigure}[b]{0.24\textwidth}
		\centering
		\includegraphics[width=\linewidth]{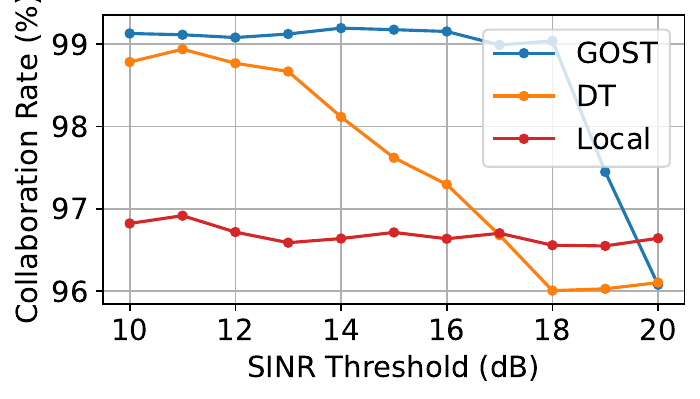}
		\caption{Collaboration rate}
		\label{track_collaboration}
	\end{subfigure}
	\hfill                     
	\begin{subfigure}[b]{0.24\textwidth}
		\centering
		\includegraphics[width=\linewidth]{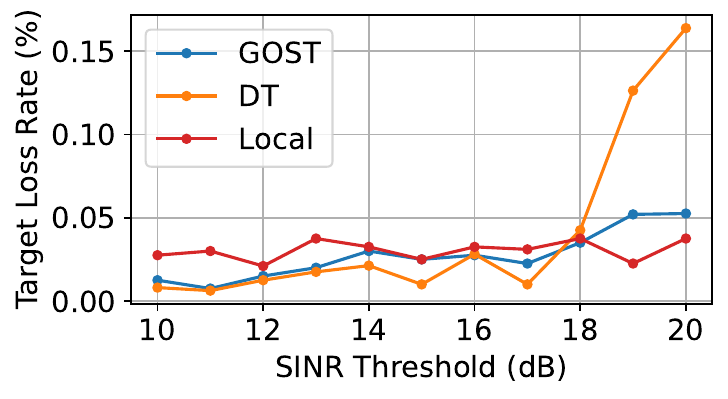}
		\caption{Target loss rate}
		\label{track_lost}
	\end{subfigure}
	
	\vspace{0.1cm}             
	
	\begin{subfigure}[b]{0.24\textwidth}
		\centering
		\includegraphics[width=\linewidth]{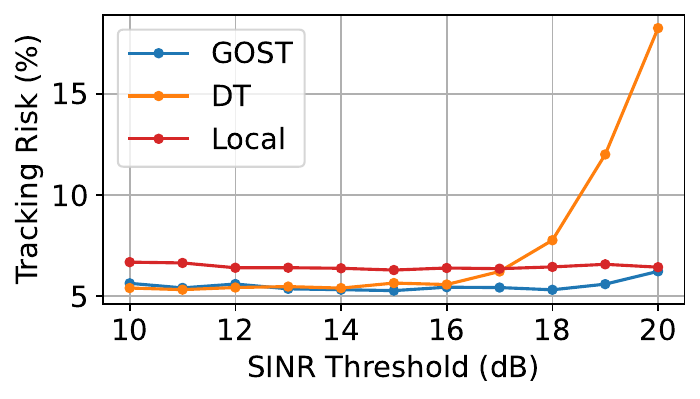}
		\caption{Tracking risk}
		\label{track_safe}
	\end{subfigure}
	\hfill
	\begin{subfigure}[b]{0.24\textwidth}
		\centering
		\includegraphics[width=\linewidth]{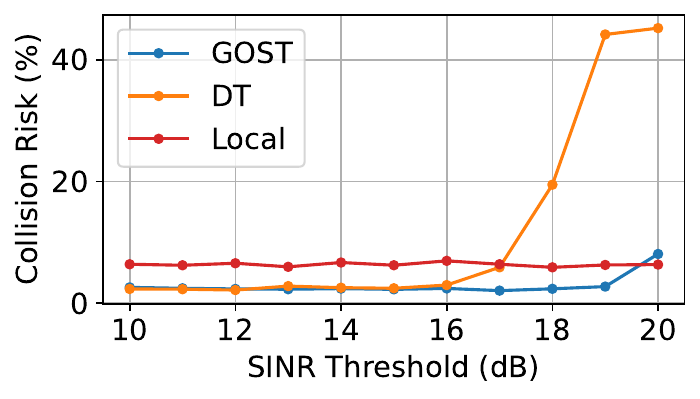}
		\caption{Collision risk}
		\label{track_collision}
	\end{subfigure}
	
	\captionsetup{subrefformat=parens}  
	\caption{Tracking performance of GOST, compressed DT, and local flight-control models under different channel conditions, evaluated by task-intent-derived indicators with different weights.}
	\label{case_track}
\end{figure}

Based on timely and semantically reliable network states, the decision agent generates flight actions targeting the optimization metrics defined in the goal-oriented principle layer. The MADDPG decision model uses a multi-objective reward that accounts for multi-cluster collaboration rate, target loss rate, tracking risk, and collision risk. Metric weights are assigned according to the priority and cost identified during intent interpretation.

This case reflects the constraint- and priority-aware treatment observed in current SAGSIN optimization studies rather than a general solution to multi-stakeholder negotiation \cite{cui2021multiobjective,tu2024priority}. Collision and minimum-distance requirements remain explicit safety constraints; target acquisition carries the highest task priority; and coverage, tracking geometry, and energy-related performance are optimized within the resulting feasible region. If communication loss makes the edge-generated command unavailable, control returns to the local controller. The example therefore helps readers connect the reviewed literature on constrained, priority-aware, and multi-objective optimization to one GOST-enabled task, while cross-operator conflict resolution remains an open topic.

Fig. \ref{case_track} compares GOST, compressed DT, and local flight-control models under varying SINR thresholds. GOST maintains the highest rewards under most channel conditions and preserves basic task performance as channel quality deteriorates. In contrast, compressed DT depends strongly on target-state estimation accuracy and its performance eventually falls below that of the local controller. When the SINR threshold exceeds 19~dB, the edge-computed GOST system may temporarily fail to generate a valid command; the declared degraded mode then transfers authority to the local decision system until link recovery. Intent parsing and multi-objective weighting balance competing metrics while prioritizing target acquisition. The target loss rate remains one to two orders of magnitude lower than the other risk indicators in the reported simulations. Sensitivity analysis of the multi-objective weights further examines the stability of this reward mechanism; the corresponding simulation study is reported in \cite{LTL_GOST_TMC}.

This case study has presented a remote multi-UAV collaborative scenario in which GOST combines task-objective interpretation with semantic-enhanced data interaction under constrained communication. The comparison against compressed DT supports GOST's timeliness, semantic fidelity, and task-utility results. Future work should examine semantic evolution under dynamic conditions and transferability across multi-task scenarios.

\section{Future Directions and Challenges}

GOST integration into SAGSIN remains an early-stage research area. This section summarizes the principal deployment bottlenecks and open questions that must be resolved before the framework can become scalable, robust, and efficient.

\subsection{Progressive Semantic Construction and Dynamic Synchronization}

SAGSIN is inherently a ``system of systems'' composed of heterogeneous assets from multiple domains (space, air, ground, and sea). A core challenge is to create a unified semantic construction process that can accurately represent diverse network entities and their complex spatiotemporal, physical, and functional relationships across multiple levels of abstraction. This requirement goes far beyond simple data-format conversion and instead demands capturing the deeper ``meaning'' of data: from orbital elements and communication payload status of satellites, to flight attitudes and sensor parameters of UAVs, to mission objectives and QoS requirements of ground vehicles. Existing ontology-based solutions ease data access but still struggle to make heterogeneous asset data truly discoverable and uniformly interpretable by domain experts\cite{Singh2020DataMF}. 



A closely related challenge concerns GOST's core capability of maintaining real-time, bidirectional data links with physical entities. Research on GOST synchronization at the semantic and data-association levels, however, remains limited. These two levels are deeply coupled with the challenges of dynamic synchronization: synchronization decisions must account for meta-knowledge such as task criticality, data volatility, and application relevance, in addition to the relationships between semantic models and the encoded entities. Future research must focus on ``semantic-driven synchronization,'' where synchronization middleware actively queries knowledge graphs to obtain real-time decision support about data flows, thereby transforming a twin from a passive data mirror into an active, intelligent information manager.

The immediate research problem is not the one-time construction of a universal SAGSIN ontology, but controlled evolution from incomplete local knowledge. Benchmarks should measure time and expert effort to reach operational coverage, mapping accuracy across organizations, constraint violations introduced by LLM proposals, incremental-update latency, and the semantic-change impact scope (i.e., the number of GOST instances affected) of an ontology revision. Open questions include how to negotiate mappings without exposing sovereign schemas, when a local concept should be promoted globally, and how running instances remain reproducible while the shared KG evolves.

\subsection{Holistic Joint Optimization of the PCCA Loop}

In SAGSIN, the lifecycle of any task can be described by the closed PCCA loop. For instance, a UAV may perceive an abnormal ground event via onboard sensors, perform preliminary computation (e.g., image recognition) using its embedded processor, communicate the results to a ground control station over a wireless link, and finally receive new commands from the station to actuate subsequent maneuvers. Isolated optimization of any single component in the PCCA chain is insufficient and can even be counterproductive. For example, maximizing communication throughput by adopting an extremely high compression ratio---thereby optimizing the communication link---may severely degrade the quality of the sensed information, leading to flawed computation and decision-making in downstream processes \cite{lu2025joint}.

The real challenge lies in treating the entire PCCA loop as an indivisible whole and performing end-to-end joint optimization to maximize ultimate task performance \cite{fan2021joint}. GOST engages with every phase of the PCCA loop and is expected to become a core enabling technology for such joint optimization.

The literature increasingly moves from scalarized network objectives toward formulations that retain service constraints, priority information, and multiple KPIs throughout optimization \cite{cui2021multiobjective,tu2024priority,lyu2024dynamic}. For the PCCA loop, an important next step is to connect these techniques across perception, communication, computing, and actuation, so that infeasible combinations are detected before execution and trade-offs remain visible. Multi-stakeholder SAGSIN further requires evidence on how operator policies, fairness, and emergency priorities interact under limited resources, especially when no jointly feasible operating point exists.



\subsection{Heterogeneous Protocol Compatibility}

SAGSIN protocol stacks are inherently heterogeneous across segments: space links follow delay-tolerant conventions, aerial segments rely on IEEE 802.11-class or 3GPP NTN air interfaces, and terrestrial and maritime segments operate their own cellular and dedicated stacks \cite{chen2025space,elmahallawy2024communication,3GPP_TR38_882}. Replacing or gateway-converting these stacks end to end would sacrifice the link-layer optimizations on which each segment depends.

Future work should therefore avoid forcing a single low-level stack across all domains, and instead develop upper-level semantic fusion interfaces that preserve local protocol efficiency while enabling cross-domain coordination. A promising direction is to position GOST as a semantic interworking layer above these heterogeneous stacks: the knowledge-based layer unifies protocol capabilities and QoS parameters in a machine-readable form \cite{karabulut2024ontologies}, the data-driven layer infers cross-segment states from heterogeneous measurements, and the goal-oriented layer selects translation points and routing strategies according to the active mission goal. Open problems include standardized semantic descriptions of protocol capabilities, consistent KPI translation across stacks, and benchmarks that quantify the interworking overhead relative to native operation \cite{cui2022sagin,du2024joint}.

\subsection{Governance for Ephemeral GOST Instances}

On-demand creation and retirement reduce resource cost but complicate accountability. Existing work provides complementary foundations rather than one complete governance solution: lifecycle reviews expose unresolved data-ownership and lifecycle gaps \cite{jones2020characterising}; ontology-based governance tracks twin-data evolution \cite{9813407}; synchronized provable possession protects distributed time-state consistency and data integrity \cite{li2022synchronized}; and formal executable models support analysis of DT behavior \cite{alhazmi2026formal}. Future research should connect these strands through compact cross-domain provenance schemas, privacy-preserving audit queries, retention policies for retired instances, and reproducible replay when models or KGs have evolved.

Trust also requires semantic-level fault containment. Future evaluations should perturb task-critical features, relations, and inference rules; measure downstream physical consequences; and test whether uncertainty-aware redundancy, constraint checks, and local fail-safe control contain the attack. These experiments are distinct from conventional packet-integrity tests because a syntactically valid message may still carry a harmful meaning.

\subsection{GOST-Empowered Internet of Agents}

In the foreseeable future, the immense complexity of configuring, managing, and optimizing SAGSIN will far exceed the cognitive limits of human operators. In SAGSIN, the ultimate user interface could be a natural language prompt, into which operators enter textual commands. The LLM, leveraging its natural language understanding, decomposes ambiguous, human-like requests into a series of structured sub-goals, constraints, and policies \cite{10726902}, and interacts with GOST to query the twin's knowledge graph for necessary contextual information. 

From the perspective of task execution in SAGSIN, each node is envisioned as an intelligent agent, forming a network of agents. GOST can integrate multi-source observational data from space, aerial, terrestrial, and maritime nodes, constructing a global environmental view that surpasses the limited perspective of any single agent. Going a step further, the virtual representation of GOST can extend to encompass the agents themselves. Given specific mission objectives, GOST can simulate and preview agent behaviors to detect potential conflicts and assess their collective contribution to the overall mission goals.

For multiple agent owners, the central challenge is governed coordination rather than unconstrained autonomy. Existing priority-aware and multi-objective studies provide tools for service differentiation and performance trade-offs \cite{cui2021multiobjective,tu2024priority}, but experimental evidence on cross-domain authentication, fairness, and conflict handling under real-time constraints remains limited. Each agent should expose capabilities, requested resources, priority information, and evidence in machine-readable form. GOST can then support pre-actuation conflict detection, policy-aware allocation of shared links and computing resources, and preservation of the reasoning trace needed to review or reproduce an outcome.



\section{Conclusion}


This survey examined twinning for SAGSIN without treating DT as a single fidelity-maximizing paradigm. DT has evolved through connected, adaptive, semantic, goal-aware, and trustworthy formulations. The remaining SAGSIN gap is compositional: large-scale intermittent operation requires task scope, semantics, resources, interaction, assurance, and lifecycle to be selected together rather than added independently.

GOST addresses that gap through an integrated six-component framework implemented across three layers. Knowledge-based semantics supports progressive cross-domain interoperability; data-driven semantics extracts and protects task-relevant state; and goal-oriented principles govern scope, multi-goal coordination, deployment, evidence, and retirement. The construction tutorial and five-dimensional evaluation framework further identify semantic robustness, constraint compliance, provenance coverage, and replayability as necessary complements to timeliness and task utility.

The multi-UAV case illustrates the operational logic under constrained communication but does not by itself establish field-wide superiority. Its role is to connect the framework's design logic to measurable synchronization and decision outcomes. Broader hardware deployments, cross-operator ontology evolution, adversarial evaluation, and multi-stakeholder governance remain open. By stating these boundaries explicitly, GOST provides a coherent research agenda for combining recent DT advances in resource-constrained SAGSIN.


\bibliographystyle{ieeetr}
\bibliography{reference_revised}

\end{document}